\documentclass[12pt,showpacs,nofootinbib]{revtex4-1}
\usepackage{multirow}
\usepackage{amssymb}
\usepackage{tikz}
\usepackage{amssymb}
\usepackage{amsmath,graphicx,dsfont,mathtools}
\usepackage{physics,tensor,cancel}
\usepackage{esvect}
\usepackage{makecell}
\usepackage[compat=1.1.0]{tikz-feynman}
\usepackage{array}
\usepackage{dcolumn}
\usepackage{braket}
\usepackage{bm}
\usepackage{enumerate}
\usepackage{slashed}
\usepackage{epstopdf}
\usepackage{amsmath}
\usepackage{dcolumn,comment,caption,booktabs}
\usepackage{slashed}
\usepackage{amsfonts}
\usepackage{adjustbox}
\usepackage{tcolorbox}
\usepackage{float}
\usepackage{collectbox}
\usepackage{listings}
\usepackage[colorlinks=true, citecolor=blue, urlcolor=blue]{hyperref}
\usepackage[mathscr]{eucal}
\usepackage{graphics}
\usepackage{subcaption}
\usepackage{natbib}

\newcommand{\be}{\begin{equation}}
	\newcommand{\ee}{\end{equation}}
\newcommand{\bea}{\begin{eqnarray}}
	\newcommand{\eea}{\end{eqnarray}}
\newcommand{\ben}{\begin{enumerate}}
	\newcommand{\een}{\end{enumerate}}
\newcommand{\bde}{\begin{widetext}}
	\newcommand{\ede}{\end{widetext}}

\newcommand{\non}{\nonumber}

\newcommand{\bc}{\begin{center}}
	\newcommand{\ec}{\end{center}}

\newcommand{\no}{\nonumber}
\newcommand{\dis}{\displaystyle}

\newcommand{\crb}[1]{{\color{blue}#1}}

\begin{document}
	
\newcommand{\AdrHEPC}{$^a$Faculty of Physics and Engineering Physics, University of Science, Ho Chi Minh City, Vietnam\\ $^b$Vietnam National University, Ho Chi Minh City, Vietnam}

\title{Hawking-Page phase transitions of black holes in the Hamiltonian formalism}
\author{Tran Ngoc Thien$^{a,b}$}
\email{tnthien@hcmus.edu.vn}
\affiliation{\AdrHEPC}
\author{Vo Quoc Phong$^{a,b}$}
\email{vqphong@hcmus.edu.vn (Corresponding author)}
\affiliation{\AdrHEPC}                    
    
\begin{abstract}
We employ the Hamiltonian formalism proposed by P. M. Dirac to investigate the Hawking-Page phase transition of the non-rotating Ba\~{n}ados-Teitelboim-Zanelli (BTZ), Reissner-Nordstr\"{o}m (RN) and Kerr-Newman (KN) black holes in both on-shell and off-shell configurations. For the BTZ black hole, the on-shell Hamiltonian agrees with the corresponding Helmholtz free energy obtained from the Euclidean action, reproducing the established Hawking-Page phase structure, while the off-shell analysis is consistent with the corresponding continuous phase transition obtained from the Euclidean approach. The same agreement is found for the on-shell RN and KN black holes. We further investigate the previously unexplored off-shell configurations of the RN and KN black holes. We find that electric charge restricts the mass range of the thermal soliton state, while rotation in the KN case removes the upper mass bound and eliminates the distinction arising from arbitrary surface contributions between the thermal soliton and black hole states. These findings uncover new aspects of the off-shell thermodynamics of charged and rotating black holes and suggest a possible connection between their phase structure and the quantum nature of gravity.
\end{abstract}
\maketitle
\tableofcontents
\break
\section{Introduction}\label{intr}

Black hole thermodynamics provides a framework for describing black holes in terms of thermodynamic quantities, such as temperature, entropy, and free energy \cite{Thermal,Thermal1,Thermal2,Thermal3}. Within this framework, black holes can exhibit thermodynamic behavior, including a variety of phase transitions and critical phenomena. In recent years, the study of black hole phase transitions has attracted considerable attention, particularly in the context of anti-de Sitter (AdS) spacetimes and extended black hole thermodynamics \cite{exThermal1,exThermal2}, where the cosmological constant is treated as a thermodynamic pressure with the corresponding thermodynamic volume as its conjugate variable. This extension enlarges the thermodynamic phase space and allows black holes to exhibit a richer variety of phase transitions and critical phenomena. In particular, the phase transition between small and large AdS black holes has been shown to share similarities with the van der Waals like phase transition \cite{vdWaals,exThermal3,exThermal4}. This analogy is also reflected in other thermodynamic characteristics: the black-hole equation of state satisfies the characteristic inflection-point conditions at the critical point, while the Gibbs free energy can exhibit a swallowtail structure associated with the coexistence of competing black-hole branches. The critical behavior is further characterized by divergent thermodynamic response functions and mean-field critical exponents. The coexistence of the small and large black-hole phases can also be characterized by the Maxwell equal-area construction \cite{Maxwell-like1}. More recently, a generalized Maxwell equal-area law has been proposed in black-hole thermodynamics, in which the standard Maxwell construction appears as a special case of a generalized free-energy landscape, providing a broader framework for describing black-hole phase transitions and their extension to the complex domain \cite{Maxwell-like2}.

Beyond the conventional van der Waals-like phase structure, more complicated thermodynamic phenomena have been identified in various gravitational theories. Reentrant phase transitions provide a qualitatively different example, in which a thermodynamic system can return to a previously encountered phase as a thermodynamic parameter is varied monotonically \cite{Reentrant1,Reentrant2}. Such behavior can involve sequences of first-order and zeroth-order phase transitions, resulting in a more intricate phase structure than that associated with a conventional transition. Recent studies have also shown that gravitational theories with higher-curvature or nonlinear interactions can exhibit multicritical phenomena, including multiple critical points and multi-phase coexistence \cite{MultiCritical1,MultiCritical2}. In addition, superfluid-like phase transitions have been reported in certain black-hole systems, where the thermodynamic phase structure exhibits analogies with superfluidity and can be characterized by distinct critical behavior \cite{Superfluid1,Superfluid2}.

More recently, attention has been extended beyond the critical point to the supercritical regime, where the distinction between different thermodynamic phases can disappear. Above the critical point, the system no longer exhibits a sharp first-order phase boundary separating the small and large black-hole phases, but instead can undergo a continuous thermodynamic crossover  \cite{Supercritical1,Supercritical2}. In particular, thermodynamic response functions may show pronounced extrema in the supercritical region, and the corresponding crossover region can be characterized by Widom lines \cite{Widom1,Widom2}. Unlike a phase boundary, a Widom line does not separate two distinct thermodynamic phases; instead, it marks a region of rapid but continuous crossover between different thermodynamic behaviors above the critical point.

Along with these developments in the real thermodynamic phase space, recent studies have also explored extensions of black-hole thermodynamics into the complex domain. In particular, the Lee-Yang framework provides a way to characterize phase transitions through the distribution of zeros of the partition function in the complex plane \cite{LeeYang1,LeeYang2}. Applications to AdS black holes have led to complex phase diagrams in which coexistence, criticality, and supercritical crossover behavior can be investigated from the analytic structure of the thermodynamic potential offering a broader framework for understanding black-hole phase transitions \cite{ComplexBH1,ComplexBH2}.

In parallel with these developments, recent research has increasingly considered the dynamical aspects of black-hole phase transitions rather than focusing exclusively on equilibrium phase diagrams. Studies of metastability, nucleation, relaxation, and critical slowing down have provided insight into how black-hole systems evolve between competing thermodynamic phases and how the characteristic timescales change near critical regions \cite{Dynamics1,Dynamics2,Dynamics3}. In particular, dynamical analyses can reveal features that are not captured by equilibrium thermodynamics alone, such as the evolution of metastable states, the kinetics of phase conversion, and the slowing down of relaxation processes near criticality. These studies indicate that black-hole phase transitions should not only be characterized by their equilibrium thermodynamic potentials, but also by the dynamical processes through which different phases emerge and evolve.

The resulting phase structure can also depend sensitively on the choice of thermodynamic ensemble. For example, fixing an extensive quantity such as electric charge or angular momentum, or instead fixing its conjugate intensive variable, can lead to different thermodynamic potentials, stability conditions, and phase structures \cite{Ensemble1,Ensemble2,Ensemble3}. Consequently, the identification of phase coexistence and thermodynamic dominance requires a consistent choice of ensemble, particularly when comparing distinct gravitational configurations. This ensemble dependence is especially relevant when different black-hole branches or different gravitational backgrounds are considered within a common thermodynamic framework.

Among the various phase transitions, the Hawking-Page phase transition \cite{Hawking-Page,HP1,HP2} is particularly important because it describes a change in thermodynamic dominance between the black hole state and the thermal soliton state within the same thermodynamic ensemble \cite{Ads1,Ads2}. Below the critical temperature, the thermal soliton state is thermodynamically favored, whereas above the critical temperature, the black hole state becomes dominant. Unlike phase transitions between different black hole states, such as the small/large black hole transition, the Hawking-Page phase transition describes a change in thermodynamic dominance between two distinct gravitational configurations rather than different states of the same black hole geometry. This feature may provide a useful setting for studying the interplay between gravitational dynamics, thermodynamics, and quantum field theory in the semiclassical description of gravity.

In this work, we employ the Hamiltonian formalism proposed by P. M. Dirac \cite{Hamilton1} to investigate the thermodynamics of the non-rotating Ba\~{n}ados-Teitelboim-Zanelli (BTZ) black hole in \crb{the} New Massive Gravity (NMG) model \cite{BTZoripaper} in both on-shell and off-shell configurations. We focus on the non-rotating sector to isolate the effects of the mass parameter and, importantly, to provide a direct comparison between the established results obtained using the Euclidean action approach \cite{BTZE=M,BTZ1,BTZ2,BTZ3} and those derived from the Hamiltonian formalism. The results show that the Hamiltonian obtained within the Dirac formalism reproduces the corresponding free energy obtained from the Euclidean action approach, providing a consistent description of the thermodynamic properties of the BTZ black hole. This agreement suggests that the Hamiltonian formalism may provide an alternative approach to deriving black hole thermodynamics and studying the Hawking-Page phase transition.

Building on the results obtained for the BTZ black hole, we further extend the Hamiltonian formalism to charged and rotating black holes. We first consider the Reissner-Nordstr\"{o}m (RN) black hole in Einstein-Maxwell theory \cite{Reissner}, which introduces the effect of electric charge into the black hole geometry. In the on-shell configuration, the free energy obtained from the Hamiltonian formalism is consistent with the corresponding result derived directly from the Euclidean action approach \cite{RN1,RN2,RN3}. This agreement demonstrates that the Hamiltonian formalism can consistently reproduce the thermodynamic free energy of the RN black hole and provides an alternative approach to studying the Hawking-Page phase transition in the presence of electric charge.

We then extend the analysis to the Kerr-Newman (KN) black hole in Einstein-Maxwell theory \cite{Kerr,Newman}, which incorporates both electric charge and angular momentum. The on-shell free energy obtained from the Hamiltonian formalism again agrees with the corresponding result derived directly from the Euclidean action approach \cite{KN1,KN2,KN3}. This consistency indicates that the Hamiltonian formalism remains applicable when both electric charge and angular momentum are present, suggesting that this approach may be extended to investigate the thermodynamics and Hawking-Page phase transition in more general black hole spacetimes.

We then extend the analysis to the off-shell configurations, where we obtain new results for the thermodynamic phase structure of charged and rotating black holes \cite{EC1,EC2,EC3}. For the RN black hole, our off-shell analysis shows that the presence of electric charge imposes both a lower and an upper bound on the mass, restricting the range in which the thermal soliton state is relevant. Nevertheless, the thermal soliton and black hole states can be considered within the same thermodynamic ensemble over this relevant range, allowing their relative thermodynamic dominance to change continuously. For the KN black hole, the combined effects of electric charge and rotation similarly impose a minimum mass threshold. However, unlike the RN case, the rotational contribution removes the upper mass bound of the thermal soliton state and eliminates the distinction arising from the arbitrary surface contributions between the thermal soliton and black hole states. Consequently, both states remain relevant throughout the thermodynamic process and can be considered within the same thermodynamic ensemble. These off-shell results provide new insights into the phase structure of charged and rotating black holes, where the continuous shift in thermodynamic dominance between competing states within a common ensemble reflects the quantum nature of the system.

Within this Hamiltonian formalism, the thermodynamic quantities of black hole systems can be derived directly, with the on-shell Hamiltonian identified with the corresponding thermodynamic Helmholtz free energy. This approach provides an independent consistency check of the results obtained from the Euclidean action. Moreover, the explicit construction of the Hamiltonian for black holes undergoing the Hawking-Page phase transition may provide a useful starting point for further exploring the quantum nature of gravity, as the Hamiltonian formulation offers a natural framework for investigating the quantization of the gravitational field.

This paper has the following structure. In Sec.~\ref{intr}, we introduce the motivation and background of the study. Sec.~\ref{sec2} presents the Hamiltonian formalism of the gravitational field proposed by Dirac, which provides the foundation for constructing the Hamiltonian of black hole systems. Sec.~\ref{secgrav} presents the gravitational action and boundary term in the New Massive Gravity model. Sec.~\ref{sec3}-\ref{sec5} apply this formalism to the BTZ, RN, KN black holes in asymptotically AdS spacetime in the on-shell background and compare the results with the reported results, before further extending the analysis to off-shell configurations. Finally, Sec.~\ref{co} summarizes the main results and discusses future directions.

\section{The Hamiltonian formalism of the gravity theory}\label{sec2}

Hamiltonian mechanics provides a reformulation of classical dynamics in terms of canonical coordinates and their conjugate momenta, based on the Legendre transformation from the Lagrangian formulation. In the Hamiltonian description, the canonical coordinates, $q$, and their conjugate momenta, $p$, are regarded as phase space variables, and the dynamics are described by the Hamiltonian density, $\mathcal{H}\left(q,p,t\right)$, rather than the Lagrangian density function, $\mathcal{L}\left(q,\dot{q},t\right)$.

For the gravitational field, the Legendre transformation associated with the Einstein-Hilbert action is singular, reflecting the diffeomorphism invariance of general relativity. Consequently, the Hamiltonian formulation of gravity is a constrained Hamiltonian system, in which the canonical variables and their conjugate momenta are subject to constraints. In this work, we apply Dirac's Hamiltonian formalism of general relativity \cite{Hamilton1}, introducing a canonical description of the gravitational field by treating the metric components $g_{\mu\nu}$ as the fundamental field variables and defining their conjugate momenta from the variation of the gravitational Lagrangian with respect to the corresponding time derivatives.

\subsection{Canonical momenta}

Greek indices $\mu,\nu,\dots$ denote spacetime coordinates $x^\mu$ and take the values $0,1,2,\dots D-1$, where $D$ is the spacetime dimension. The canonical momenta $p^{\mu \nu}$ conjugate to the metric variables $g_{\mu\nu}$ are defined as the coefficients of the variations of their corresponding time derivatives, $\delta(\partial_0 g_{\mu\nu})$, in the variation of the gravitational Lagrangian:
\begin{eqnarray}
\int \delta \mathcal{L}d^3x = \int p^{\mu\nu}\delta (\partial_0 g_{\mu\nu })d^3x,
\end{eqnarray}

Since the Legendre transformation is singular, the relations between the canonical momenta and the time derivatives of the metric components cannot be inverted uniquely to determine all the corresponding velocities. In particular, the components  $p^{\mu 0}$ vanish identically and therefore give rise to the primary constraints
\begin{eqnarray}
p^{\mu 0} \approx 0.
\end{eqnarray}

To facilitate the identification of the non-vanishing canonical momenta, we introduce an equivalent Lagrangian $\mathcal{L}^*$ by adding a total divergence to the original gravitational Lagrangian $\mathcal{L}$, such that the bulk equations of motion remain unchanged. Explicitly, we define
\begin{eqnarray}
\mathcal{L}^* - \mathcal{L}=
\partial_v (\sqrt{-g}g^{00})
\partial_0
\left(
\frac{g^{v0}}{g^{00}}
\right)
-
\partial_0 (\sqrt{-g}g^{00})
\partial_v
\left(
\frac{g^{v0}}{g^{00}}
\right).
\end{eqnarray}

This modification changes the dependence of the Lagrangian on the time derivatives of the metric components, in particular on $\partial_0 g_{rs}$, and is therefore useful for identifying the non-vanishing canonical momenta $p^{rs}$ within the constrained Hamiltonian formulation.

To express the variation of $\mathcal{L}^*$ in terms of the spatial components of the metric, we introduce the symmetric auxiliary quantity \cite{Hamilton1}:
\begin{eqnarray}
e^{\mu\nu} = e^{\nu\mu} = g^{\mu\nu} - \frac{g^{\mu 0}g^{\nu 0}}{g^{00}},
\end{eqnarray}

Setting $\mu = 0$, we obtain

\begin{eqnarray}
e^{0\nu} = e^{\nu 0} = g^{0\nu} - \frac{g^{0 0}g^{\nu 0}}{g^{00}} = g^{0\nu} - g^{\nu 0} = 0.
\end{eqnarray}

The division by $g^{00}$ is well-defined because $g^{00}$ is a nonzero component of the inverse metric $g^{\mu\nu}$. Therefore $e^{\mu\nu}$ vanishes if either $\mu$ or $\nu$ is zero. The Latin indices $r,s,\dots,$ denote spatial coordinates and take the values $1,2,\dots, D-1$. We consider

\begin{eqnarray}
e^{r\mu}g_{\mu\nu}&=&e^{r 0}g_{0\nu} + e^{r s}g_{s\nu} \no\\
&=&
g^{r 0}g_{0\nu} - \frac{g^{r 0}g^{0 0}}{g^{00}}g_{0\nu} + 
g^{rs}g_{s\nu}  - \frac{g^{r 0}g^{s 0}}{g^{00}}g_{s\nu} \no\\
&=&
g^{r 0}g_{0\nu} - g^{r 0}g_{0\nu} + 
g^{rs}g_{s\nu}  - \frac{g^{r 0}g^{s 0}}{g^{00}}g_{s\nu} \no\\
&=&
\left(g^{rs}  - \frac{g^{r 0}g^{s 0}}{g^{00}}\right)g_{s\nu} = e^{rs}g_{s\nu},
\end{eqnarray}

the division by $g^{00}$ is well-defined because $g^{00}\neq0$. Here, $g^{00}$ and $g^{r0}$ denote specific components of the inverse metric $g^{\mu\nu}$, while $g_{0\nu}$ denotes a component of the covariant metric $g_{\mu\nu}$. The index $0$ in these components is fixed and is not subject to summation. This should be distinguished from an expression such as $g^{0\lambda}g_{\lambda\nu}$, in which the repeated index $\lambda$ is summed according to the Einstein summation convention. Likewise, the repeated spatial index $s$ is summed over, with $s=1,\dots,D-1$. We can separate the calculation into two cases, corresponding to $\nu=0$ and $\nu=b$, respectively. For $\nu=0$, we have

\begin{eqnarray}
e^{r\mu}g_{\mu 0}=e^{rs}g_{s 0}
=g^{rs}g_{s 0}  - \frac{g^{r 0}g^{s 0}}{g^{00}}g_{s0}
=\left(g^{rs}g_{s 0}\right)  - \frac{g^{r 0}\left(g^{s 0}g_{s0}\right)}{g^{00}}
=g^r_0 -\frac{g^{r0}g^0_0}{g^{00}} = -\frac{g^{r0}}{g^{00}}.
\end{eqnarray}

Therefore, for $\nu= b \neq 0$, we have

\begin{eqnarray}
e^{r\mu}g_{\mu b}=e^{rs}g_{s b}
=g^{rs}g_{s b}  - \frac{g^{r 0}g^{s 0}}{g^{00}}g_{sb}
=\left(g^{rs}g_{s b}\right)  - \frac{g^{r 0}\left(g^{s 0}g_{sb}\right)}{g^{00}}
=g^r_b -\frac{g^{r0}g^0_b}{g^{00}} = g^r_b,
\end{eqnarray}

Thus, in spatial coordinates

\begin{eqnarray}
e^{rs}g_{sb} = g^r_b,
\end{eqnarray}

hence $e^{rs}$ is the reciprocal of the spatial metric $g_{rs}$ and can be regarded as the fundamental tensor associated with the spatial hypersurface of constant $x^0$. Thus, the variation of the modified Lagrangian can be expressed as
\begin{eqnarray}
\delta \mathcal{L}^*&=&2\sqrt{-g}(e^{ra}e^{sb}-e^{rs}e^{ab})\Gamma^0_{ab}\delta \Gamma^0_{rs}/g^{00}\no\\
&=&-\sqrt{-g}(e^{ra}e^{sb}-e^{rs}e^{ab})\Gamma^0_{ab}\delta( \partial_0 g_{rs}).
\end{eqnarray}
Accordingly, the non-vanishing components of the canonical momenta conjugate to the metric variables are given by
\begin{eqnarray}
p^{rs}=\sqrt{-g}(e^{rs}e^{ab}-e^{ra}e^{sb})\Gamma^0_{ab}.\label{p}
\end{eqnarray}

\subsection{The Hamiltonian of the gravitational field}

Having identified the canonical field variables and their corresponding canonical momenta, we now construct the canonical Hamiltonian of the gravitational field through the Legendre transformation. In the canonical formulation, the metric components $g_{\mu\nu}$ serve as the fundamental field variables, while their time derivatives $\partial_0 g_{\mu\nu}$ play the role of the corresponding generalized velocities. The canonical Hamiltonian is therefore obtained from the Lagrangian density as
\begin{eqnarray}
H \equiv H_c = \int \left(p^{rs}\partial_0 g_{rs} -\mathcal{L}\right)d^3x.
\end{eqnarray}

Thus, the Dirac Hamiltonian formalism provides the canonical framework for constructing the Hamiltonian, while the explicit form of the Lagrangian is determined by the underlying gravitational theory. 

In this work, we employ the radial ADM-like decomposition \cite{ADM1,ADM2} of the metric to construct the corresponding gravitational actions and boundary terms, from which the explicit Lagrangian densities are obtained. We then apply the Dirac Hamiltonian formalism to investigate the Hawking-Page phase transition in AdS spacetimes for the BTZ, RN, and KN black holes. For the BTZ black hole, we employ the established action and boundary terms of the New Massive Gravity (NMG) model, while the RN and KN black holes are considered within Einstein-Maxwell theory. We then compare our results with previous results obtained directly from the Euclidean action approach.

\section{The Gravitational action and boundary term in the New Massive Gravity}\label{secgrav}

For the BTZ black hole, we employ the three-dimensional spacetime, where the spacetime coordinates are $x^\mu = (t,r,\phi)$, New Massive Gravity (NMG) model to study the effects of higher-curvature corrections on the Hawking-Page phase transition. The Einstein-Hilbert action is given by
\begin{eqnarray}
I_{\text{EH}}&=&\frac{1}{16\pi G}\int_\mathcal{M}d^3x\sqrt{-g}(R-2\lambda),
\end{eqnarray}
and the higher-derivative terms in the absence of a matter source. In three-dimensional spacetime, the quadratic-curvature sector can be constructed from $R_{\mu\nu}R^{\mu\nu}$ and $R^2$ \cite{HigherCurvature}. 

\begin{eqnarray}
I_{\text{HC}}&=&-\frac{1}{16\pi Gm^2}\int_\mathcal{M}d^3x\sqrt{-g}(R_{\mu\nu}R^{\mu\nu}-\frac{3}{8}R^2),
\end{eqnarray}
where, $m$ is the mass parameter controlling the strength of the higher-curvature corrections, and the Einstein-gravity limit is recovered as $m \rightarrow \infty$ \cite{Gravital-Action,BTZ-Metric0, BTZ-Metric1}. The combination $R_{\mu\nu}R^{\mu\nu}-\dis\frac{3}{8}R^2$ defines the higher-curvature sector of NMG and is chosen such that the linearized theory describes the desired massive spin-2 degrees of freedom, being equivalent to the Fierz-Pauli theory at the linearized level \cite{HigherCurvature}.

The equations of motion \cite{BTZ-Metric0, BTZ-Metric1} derived from the bulk action, $I_{\text{bulk}} = I_{\text{EH}} + I_{\text{HC}}$ are given by
\begin{eqnarray}
G_{\mu\nu} - \lambda g_{\mu\nu} - \frac{1}{2m^2}Y_{\mu\nu} = 0,
\end{eqnarray}
where $G_{\mu\nu}$ is the Einstein tensor, $G_{\mu\nu} = R_{\mu\nu} - \dis \frac{1}{2}g_{\mu\nu}R$, and the higher-curvature term $Y_{\mu\nu}$,
\begin{eqnarray}
Y_{\mu\nu}&= 2 \nabla^2 R_{\mu\nu} - \frac{1}{2}\left(\nabla_\mu \nabla_\nu R + g_{\mu\nu} \nabla^2 R \right)-8R^{\sigma}_{\mu}R_{\nu \sigma}\nonumber\\
&+\frac{9}{2}R R_{\mu\nu} + g_{\mu\nu} \left[ 3R^{\sigma \rho}R_{\sigma \rho} - \frac{13}{8}R^2\right].
\end{eqnarray}

However, to be able to use the principle of least action to get the equations of motion \cite{GHY-term}, we introduce an auxiliary field:
\begin{eqnarray}
f_{\mu\nu}=-\frac{2}{m^2}
\left(
R_{\mu\nu}-\frac{1}{4}Rg_{\mu\nu}
\right),
\end{eqnarray}
which is the solution obtained from the bulk action in its second-order form,
\begin{eqnarray}
I_{\text{bulk}}&=&\frac{1}{16\pi G}\int_\mathcal{M}d^3x\sqrt{-g}\left[
R-2\lambda 
+f^{\mu\nu}G_{\mu\nu} + \frac{1}{4}m^2 \left(
f^{\mu\nu}f_{\mu\nu} - f^2 \right)
\right],
\end{eqnarray}
with $f = g^{\mu\nu}f_{\mu\nu}$. The variation of the higher-derivative NMG action\crb{,} after varying the bulk equations of motion, is
\begin{eqnarray}
\delta S_{NMG} = \frac{1}{16\pi G}\int_\mathcal{M}d^3x \sqrt{-g}D_\mu v^\mu 
\end{eqnarray}
where the non-vanishing boundary contribution is given by
\begin{eqnarray}
v^\mu &=& \left(1-\frac{1}{2}f \right)\left[g^{\mu\nu}D^\rho(\delta g_{\nu\rho}) - g^{\rho\sigma}D^\mu(\delta g_{\rho\sigma})\right]\non\\
&&+f^{\rho\sigma}g^{\mu\nu}D_\rho (\delta g_{\sigma \nu})
-\frac{1}{2}f^{\mu\nu}g^{\rho\sigma}D_\nu (\delta g_{\rho\sigma})
-\frac{1}{2}f^{\rho\sigma}D^\mu (\delta g_{\rho\sigma}).
\end{eqnarray}

To eliminate these terms, we add local boundary terms that generalize the Gibbons-Hawking term. With the spacetime coordinates, $x^\mu = (t,r,\phi)$, we decompose the metric of $\mathcal{M}$ with respect to the boundary hypersurface $\partial \mathcal{M}$, defined by $r=const$, where $t$ and $\phi$ are the intrinsic coordinates on $\partial\mathcal{M}$, and $i,j$ label these intrinsic coordinates. 
\begin{eqnarray}
ds^2&=&N^2dr^2 + \gamma_{ij}(dx^i + N^idr)(dx^j + N^jdr),
\end{eqnarray}
where $\gamma_{ij}$ defines the induced metric on the boundary $\partial \mathcal{M}$, while $N$ denotes the lapse function and $N^i$ is the shift vector. Using Stokes' theorem, the integral of the total derivative over $\mathcal{M}$ can be converted into a boundary integral over $\partial \mathcal{M}$,
\begin{eqnarray}
\delta S_{NMG} = \frac{1}{16\pi G}\int_{\partial \mathcal{M}}d^2x \sqrt{-\gamma}n_\mu v^\mu, \hspace{10mm} n_\mu = (0, N, 0),
\end{eqnarray}
where $n^\mu$ is the normal vector to a hypersurface of constant $r$. The three-dimensional geometry is encoded in the intrinsic boundary determined by the boundary metric and the extrinsic curvature tensor in terms of the normal vectors,
\begin{eqnarray}
K_{\mu\nu} = -\frac{1}{2}\left( D_\mu n_\nu + D_\nu n_\mu \right).
\end{eqnarray}

The relevant boundary components read
\begin{eqnarray}
K_{ij} = \Gamma^r_{ij}N = -\frac{1}{2N}\left(\partial_r \gamma_{ij} -\nabla_i N_j -\nabla_j N_i \right).
\end{eqnarray}

Next, we express the radial component $v^r$ and need to decompose the auxiliary $f^{\mu\nu}$,
\begin{eqnarray}
f^{\mu\nu} = \begin{pmatrix}
f^{rr} & f^{ri}\\
f^{rj} & f^{ij}
\end{pmatrix}.
\end{eqnarray}

We redefine $f = \gamma^{ij} f_{ij}$,
\begin{eqnarray}
\delta S_{NMG} &=& \frac{1}{16\pi G}\int_{\partial \mathcal{M}}d^2x \sqrt{-\gamma}N v^r \non\\
&=& \frac{1}{16\pi G}\int_{\partial \mathcal{M}}d^2x \sqrt{-\gamma}\left(
2\gamma^{ij} + f^{ij} - f \gamma^{ij}
\right)\delta K_{ij},
\end{eqnarray}

After integrating by parts the terms proportional to $f^{rk}\gamma^{ij}\nabla_k (\delta \gamma_{ij})$, we obatin the generalized Gibbons-Hawking term,
\begin{eqnarray}
I_{\text{GGH}}&=&\frac{1}{16\pi G}\int_{\partial\mathcal{M}}d^2x\sqrt{-\gamma}\left(
-2K-\hat{f}^{ij}K_{ij}+\hat{f}K
\right),
\end{eqnarray}
where
\begin{eqnarray}
\hat{f}^{ij} &=& f^{ij}+2\left(f^{ri}N^j - f^{rj}N^i\right) +f^{rr}N^iN^j,\\
K &=& \gamma^{ij}K_{ij}, \hspace{10mm}\hat{f} = \gamma_{ij}\hat{f}^{ij},
\end{eqnarray}

\section{Ba\~{n}ados-Teitelboim-Zanelli black hole}\label{sec3}

The metric of the non-rotating Ba\~{n}ados-Teitelboim-Zanelli black hole in the three-dimensional spacetime system in AdS \cite{BTZ1,BTZ2,BTZ3}, where the spacetime coordinates are $x^\mu=(t,r,\phi)$, with the spacetime index $\mu=0,1,2$,  has the form 
\begin{eqnarray}
ds^2_{BTZ}=-f(r)dt^2 + \frac{1}{f(r)}dr^2+r^2d\phi^2,
\end{eqnarray}
with $f(r)=-M+\dis{\frac{r^2}{l^2}}$. In which, the integration constant $M$ corresponds to the ADM mass and $l$ is the AdS radius. The event horizon, $r_H = l\sqrt{M}$, and the temperature as $T_H=\dis\frac{1}{4\pi}\left.\dis\frac{df}{dr}\right|_{r_H}=\dis\frac{\sqrt{M}}{2\pi l}.$

To study the Hawking-Page phase transition using Dirac Hamiltonian formalism, an explicit gravitational Lagrangian is required. We use 3D NMG model for the BTZ black hole to investigate the effects of higher-curvature corrections on the Hawking-Page phase transition. This provides a well-established starting point for obtaining the corresponding Lagrangian density and allows the results to be checked against known properties of NMG black holes \cite{BTZ1,BTZ2,BTZ3}.

From the bulk action, we add the generalized Gibbons-Hawking term, $I_{\text{GGH}}$, and to be able to get a finite Euclidean action, we must add a term $I_{\text{ct}}$ that normalizes the action. The total action is now of the form \cite{ct1,ct2,ct3,ct4}
\begin{eqnarray}
I&=&I_{\text{bulk}}+I_{\text{GGH}}+I_{\text{ct}},\\
I_{\text{GGH}}&=&\frac{1}{16\pi G}\int_{\partial\mathcal{M}}d^2x\sqrt{-\gamma}\left(
-2K-\hat{f}^{ij}K_{ij}+\hat{f}K
\right),\\
I_{\text{ct}}&=&-\frac{1}{8\pi G l}\left(
1-\frac{1}{2m^2 l^2}
\right)\int_{\partial\mathcal{M}}d^2x\sqrt{-\gamma},
\end{eqnarray}

From the action established above, the Lagrangian is decomposed into a bulk contribution $\mathcal{L}_{\mathcal{M}}$ and a boundary contribution $\mathcal{L}_{\partial\mathcal{M}}$. Accordingly, the action takes the form

\begin{eqnarray}
I=\int_\mathcal{M}d^3x\hspace{1mm} \mathcal{L}_\mathcal{M} + \int_{\partial\mathcal{M}}d^2x\hspace{1mm} \mathcal{L}_{\partial\mathcal{M}},
\end{eqnarray}
with
\begin{eqnarray}
\mathcal{L}_\mathcal{M} &=&\frac{1}{16\pi G}\sqrt{-g}\left(R-2\lambda-\frac{1}{m^2}R_{\mu\nu}R^{\mu\nu}+\frac{3}{8m^2}R^2\right).\\
\mathcal{L}_{\partial\mathcal{M}}&=&\frac{1}{16\pi G}\sqrt{-\gamma}\left[
-2K-\hat{f}^{ij}K_{ij}+\hat{f}K
-\frac{2}{l}\left(
1-\frac{1}{2m^2 l^2}
\right)
\right].
\end{eqnarray}

Thus the corresponding Hamiltonian is then decomposed into the bulk and boundary contributions. Since the metric is time independent and contains no $dtdx^a$ cross terms, we have $\partial_0 g_{\mu\nu} = 0$ and $g_{0a} = 0$, which imply $\Gamma^0_{ab} = 0$. Therefore, $p^{rs}$ vanishes and hence $p^{rs}\partial_0 g_{rs} = 0$

\begin{eqnarray}
H_{\mathcal{M}}&=\dis\int_{\mathcal{M}} d^2x\hspace{1mm} (p^{rs}_{_{_{\mathcal{M}}}}\partial_0 g_{rs_{\mathcal{M}}}-\mathcal{L}_\mathcal{M})=&-\int_{\mathcal{M}} d^2x\hspace{1mm} \mathcal{L}_\mathcal{M},\\
H_{\partial\mathcal{M}}&=\dis\int_{\partial\mathcal{M}} dx\hspace{1mm} (p^{rs}_{_{_{\partial\mathcal{M}}}}\partial_0 g_{rs_{\partial\mathcal{M}}}-\mathcal{L}_{\partial\mathcal{M}})=&-\int_{\partial\mathcal{M}} dx\hspace{1mm} \mathcal{L}_{\partial\mathcal{M}}.
\end{eqnarray}

\subsection{The Hamiltonian of the black hole}

The metric derived from $ds^2_{BTZ}$ in the space $\mathcal{M}$ has the form
\begin{eqnarray}
g_{\mu\nu}=\begin{pmatrix}
-f(r) 	&0 					&0\\
0		&\frac{1}{f(r)}		&0\\
0		&0					&r^2
\end{pmatrix},\hspace{10mm}g^{\mu\nu}=\begin{pmatrix}
-\frac{1}{f(r)} 		&0 					&0\\
0						&f(r)				&0\\
0						&0					&\dis\frac{1}{r^2}
\end{pmatrix}.
\end{eqnarray}

We can establish the Ricci tensor, $R_{\mu\nu}$, from the non-zero component,

\begin{eqnarray}
R_{\mu\nu} = \begin{pmatrix}
R_{00} & 0 & 0\\
0 &R_{11} &0\\
0 &0 &R_{22}
\end{pmatrix}= \begin{pmatrix}
\dis\frac{2}{l^2}f & 0 & 0\\
0 & -\dis\frac{2}{l^2} \frac{1}{f} &0\\
0 &0 &-\dis\frac{2}{l^2} r^2
\end{pmatrix} = -\frac{2}{l^2}g_{\mu\nu},
\end{eqnarray}

and $\lambda=-\dis\frac{1}{l^2}-\dis\frac{1}{4m^2l^4}$, we obtain the gravitational Lagrangian on the $\mathcal{M}$ surface:
\begin{eqnarray}
\mathcal{L}_\mathcal{M}
&=&\frac{r}{16\pi G}
\left(
-\frac{4}{l^2}
+\frac{2}{m^2l^4}
\right).
\end{eqnarray}

We decompose the BTZ metric on $\mathcal{M}$ with respect to the boundary hypersurface $\partial\mathcal{M}$, defined by $r=\mathrm{const}$, with the spacetime coordinates $x^\mu=(t,r,\phi)$, where $t$ and $\phi$ are the intrinsic coordinates on $\partial\mathcal{M}$, labeled by $i,j$

\begin{eqnarray}
ds^2_{BTZ}&=&N^2dr^2 + \gamma_{ij}(dx^i + N^idr)(dx^j + N^jdr),
\end{eqnarray}

Since there are no cross terms of the form $dx^i dr$, the shift vector, $N^i = 0$, the lapse function $N$, and the induced metric on the boundary $\partial \mathcal{M}$, $\gamma_{ij}$, are
\begin{eqnarray}
N = \dis\sqrt{\frac{1}{f(r)}}, \hspace{10mm}
\gamma_{ij}= \dis \begin{pmatrix}
-f(r)		&0\\
0		&r^2
\end{pmatrix}.
\end{eqnarray}

The auxiliary field has the form
\begin{eqnarray}
f_{\mu\nu} = -\frac{2}{m^2}\left(R_{\mu\nu} - \frac{1}{4}Rg_{\mu\nu}\right) = \frac{1}{m^2 l^2}g_{\mu\nu},
\end{eqnarray}

Its form on the boundary $\partial \mathcal{M}$ is therefore
\begin{eqnarray}
f_{ij} = \frac{1}{m^2 l^2}g_{ij} = \frac{1}{m^2 l^2}\gamma_{ij} ,
\end{eqnarray}

with $\hat{f}^{ij}=f^{ij}=\dis\frac{1}{m^2l^2}\gamma^{ij}$ and the extrinsic curvature, $K_{ij}=-\dis\frac{\sqrt{f(r)}}{2}\partial_r \gamma_{ij}$. The Lagrangian on the hypersurface $\partial \mathcal{M}$ is then
\begin{eqnarray}
\mathcal{L}_{\partial\mathcal{M}}&=&\frac{1}{l^2}\frac{1}{16\pi G}\left(
1-\frac{1}{2m^2 l^2}
\right)\left[\left(\sqrt{r^2 - r_H^2} - r \right)^2 +\left(r^2 - r_H^2\right) + r^2 \right].
\end{eqnarray}

We derive the Hamiltonian contributions from the spatial metric and its boundary induced metric,
\begin{eqnarray}
H_\mathcal{M}&=&\frac{1}{4 Gl^2}
\left(
1
-\frac{1}{2m^2l^2}
\right)\left(r^2 - r_H^2\right),\\
\no\\
H_{\partial\mathcal{M}}&=&-\frac{1}{8 Gl^2}\left(
1-\frac{1}{2m^2 l^2}
\right)\left[\left(\sqrt{r^2 - r_H^2} - r \right)^2 +\left(r^2 - r_H^2\right) + r^2 \right].
\end{eqnarray}

Therefore, the Hamiltonian of the BTZ black hole has the form
\begin{eqnarray}
H_{bh} &=& H_\mathcal{M} + H_{\partial\mathcal{M}}
= \frac{1}{8 Gl^2}
\left(
1
-\frac{1}{2m^2l^2}
\right) \left[
2r^2 - 2r^2_H -2r^2 +r^2_H - r^2 \left(\sqrt{1-\frac{r^2_H}{r^2}}-1 \right)^2
\right]
\no\\
&=&-\frac{M}{8 G}\left(
1-\frac{1}{2m^2l^2 }
\right)+\mathcal{O}\left(\frac{1}{r^2} \right),
\end{eqnarray}

where $M = \dis\frac{r_H^2}{l^2}$ and  $r^2 \left(\sqrt{1-\dis\frac{r^2_H}{r^2}}-1 \right)^2 \rightarrow 0 $ when $r \rightarrow \infty$. Thus, the Hamiltonian obtained independently from the Hamiltonian formalism is
\begin{eqnarray}
H_{bh}=-\frac{M}{8 G}\left(
1-\frac{1}{2m^2l^2 }
\right)+\mathcal{O}\left(\frac{1}{r^2} \right).
\end{eqnarray}

On the other hand, the Helmholtz free energy obtained directly from the Euclidean action \cite{BTZ1,BTZ2,BTZ3} is given by 
\begin{eqnarray}
F_{bh}=-\frac{M}{8 G}\left(
1-\frac{1}{2m^2l^2 }
\right)+\mathcal{O}\left(\frac{1}{r^2} \right).
\end{eqnarray}

Therefore, we find that the Hamiltonian obtained from the Hamiltonian formalism agrees with the Helmholtz free energy obtained independently from the Euclidean action. This agreement demonstrates that the Hamiltonian formalism provides an alternative approach for obtaining the thermodynamic free energy of the BTZ black hole.

\subsection{On-Shell}

We investigate the phase transition by calculating the Hamiltonian (free energy) of the soliton state. Here, we set $M=-1$ in the BTZ metric to specify the AdS vacuum geometry used to evaluate $H_{soliton}$. In this case, the equation $f(r)=0$ has no real solution, which indicates the absence of a real event horizon. Thus, the Hamiltonian of the soliton state is
\begin{eqnarray}
F_{soliton}=H_{soliton}&=&H_{\mathcal{M}_{soliton}} + H_{\partial \mathcal{M}_{soliton}}\no\\
&=&-\frac{1}{8G}\left(
1-\frac{1}{2m^2l^2}
\right).
\end{eqnarray}

The phase transition occurs when the free energies of the two phases are equal $F_{bh}=F_{soliton}$ or equivalently, $H_{bh}=H_{soliton}$ in our formulation. This condition determines the critical value of the BTZ mass parameter, $M=1$, and hence the critical Hawking temperature $T_C=\dis\frac{1}{2\pi l}$.

When the temperature is below the critical temperature, $T< T_C$, the soliton phase is thermodynamically favored, conversely, when the temperature is above the critical temperature, $T> T_C$, the black hole phase is thermodynamically favored. We can reformulate the black hole free energy, $F_{bh}$, in terms of the BTZ black hole Hawking temperature, $T =T_H=\dis\frac{\sqrt{M}}{\dis 2\pi l}$.

\begin{eqnarray}
F_{bh}=-\frac{4 \pi^2 l^2 T^2}{8 G}\left(
1-\frac{1}{2m^2l^2 }
\right).
\end{eqnarray}

\begin{figure}[h!]
\centering
\includegraphics[width=\textwidth]{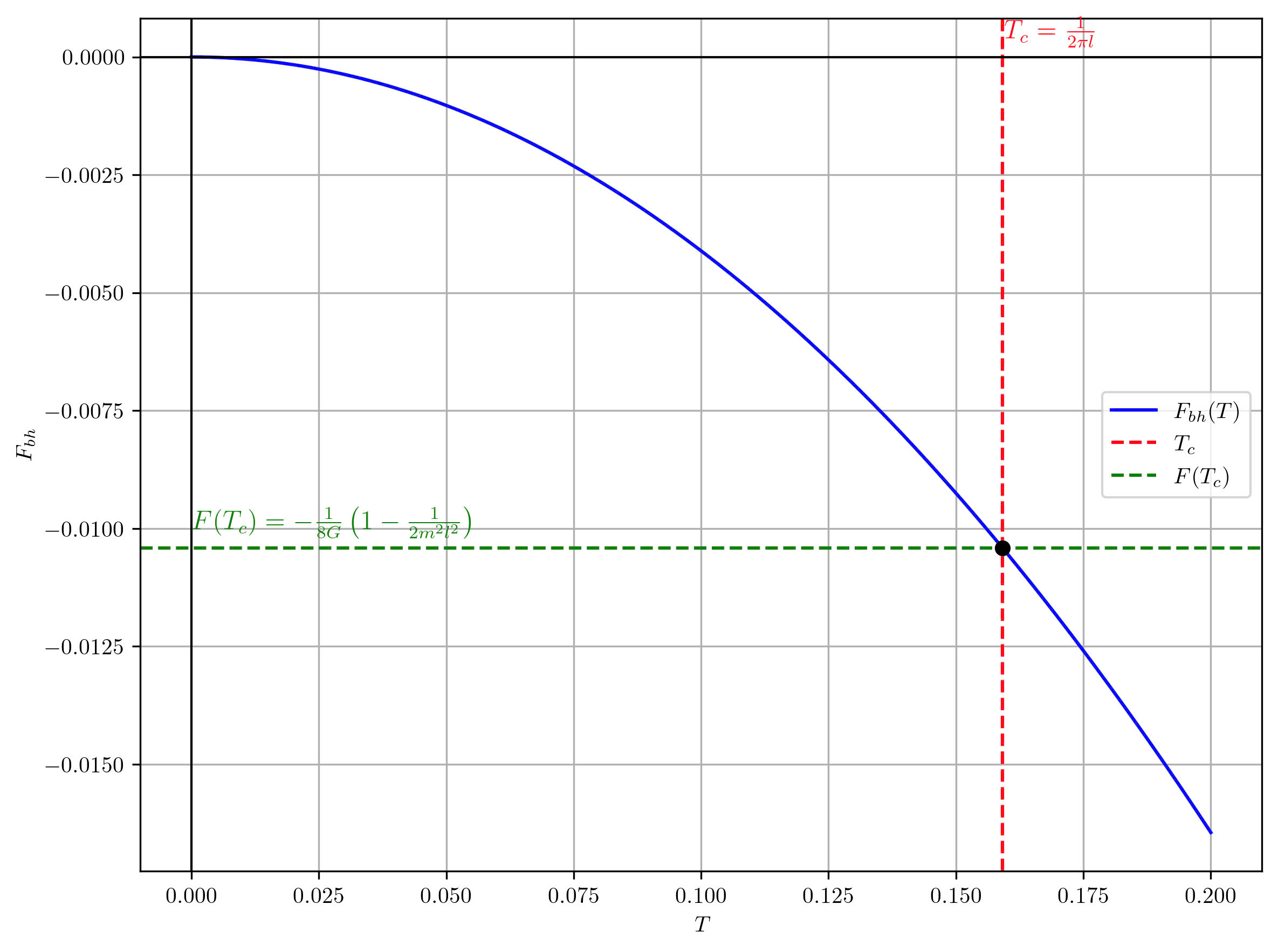}
\caption{The on-shell free energy as a function of temperature for the BTZ black hole.}\label{1}
\end{figure}

Fig.\ref{1} indicates that the phase transition occurs at the critical temperature, $T_C$, from the thermal soliton state to the black hole state. Since the phase transition occurs at a single point, in the on-shell limit we can only assume that at the critical temperature $T_C$ the system instantaneously transitions from the thermal soliton state to the black hole state. We can conclude that the Hawking-Page phase transition in this case is a first-order phase transition.

\subsection{Off-Shell}

We consider the off-shell configuration, which allows the black hole system to deviate from thermal equilibrium and can therefore be interpreted as describing thermal fluctuations around the equilibrium state. Such a deviation is characterized by the dimensionless ratio $\alpha = \frac{\beta_H}{\beta}$ and represented by a conical geometry $\mathcal{M}_\alpha$ with a deficit angle $2\pi(1-\alpha)$ and the singular set $\Sigma$. When the equilibrium $\alpha = 1$ is satisfied, the corresponding on-shell thermodynamic quantities are recovered. Therefore the off-shell analysis provides a way to examine how the thermodynamic properties evolve away from equilibrium. The Ricci scalar and Ricci tensor now take the form
\begin{eqnarray}
R_{\mu\nu}^{(\alpha)}&=& R_{\mu\nu} + 2\pi(1-\alpha)\left(n_\mu n_\nu \right)\delta_\Sigma ,\\
R^{(\alpha)}&=&R + 4\pi(1-\alpha)\delta_\Sigma,
\end{eqnarray}
with $\delta_\Sigma$ being the delta function at the singular point $\dis\int_{\mathcal{M}}f\delta_\Sigma =\int_\Sigma f$, and $n^s=n^s_\mu dx^\mu $ $(s=1,2)$ being two vectors normal and orthogonal to the singular point set $\Sigma$ where $(n_\mu n_\nu)=\dis\sum^{2}_{s=1}n^s_\mu n^s_\nu$. We aplly the off-shell condition is applied to the Lagrangian as follows
\begin{eqnarray}
\mathcal{L}_\mathcal{M} &=&\frac{1}{16\pi G}\sqrt{-g}\left[R^{(\alpha)}-2\lambda-\frac{1}{m^2}R^{(\alpha)}_{\mu\nu}R^{(\alpha),\mu\nu}+\frac{3}{8m^2}\left(R^{(\alpha)}\right)^2\right]\\
&=&\hspace{3mm}\frac{1}{16\pi G}\sqrt{-g}\left(R -2\lambda-\frac{1}{m^2}R_{\mu\nu}R^{\mu\nu}+\frac{3}{8m^2}R^2\right)\no\\
&&+\frac{1}{16\pi G}\sqrt{-g}\left\{
 4\pi(1-\alpha)\delta_\Sigma \right.\no\\
&&\hspace{3mm}-\frac{1}{m^2}
\left[ 2\pi(1-\alpha)R^{\mu\nu}\left(n_\mu n_\nu \right)\delta_\Sigma\right]
-\frac{1}{m^2}
\left[2\pi(1-\alpha)R_{\mu\nu}\left(n^\mu n^\nu \right)\delta_\Sigma\right]\no\\
&&\left.\hspace{3mm}-\frac{1}{m^2}4\pi^2 (1-\alpha)^2(n^\mu n^\nu)(n_\mu n_\nu)\delta^2_\Sigma +\frac{3}{8m^2}\left[ 8\pi R(1-\alpha)\delta_\Sigma +
\left(4\pi(1-\alpha)\delta_\Sigma\right)^2\right]\right\}\no\\
&=&\mathcal{L}_{\mathcal{M}_{\text{bulk}}} + \mathcal{L}_{\mathcal{M}_{\text{singular}}}.
\end{eqnarray}

Considering only $H_{\mathcal{M}_{\text{singular}}}$, we separate the terms according to $\delta_\Sigma$ and $\delta^2_{\Sigma}$, so that
\begin{eqnarray}
\mathcal{L}_{\mathcal{M}_{\text{singular}}}&=&
\hspace{3mm}\frac{1}{16\pi G}\sqrt{-g}\delta_\Sigma\left\{
4\pi(1-\alpha) +\frac{3}{8m^2}\left[ 8\pi R(1-\alpha)\right] \right.\no\\
&&\hspace{5mm}\left.-\frac{1}{m^2}
\left[ 2\pi(1-\alpha)R^{\mu\nu}\left(n_\mu n_\nu \right)\right]
-\frac{1}{m^2}
\left[2\pi(1-\alpha)R_{\mu\nu}\left(n^\mu n^\nu \right)\right]\right\}\no\\
&&+\frac{1}{16\pi G}\sqrt{-g}\delta^2_\Sigma
\left[
-\frac{1}{m^2}4\pi^2 (1-\alpha)^2(n^\mu n^\nu)(n_\mu n_\nu)
+\frac{1}{m^2}6\pi^2(1-\alpha)^2
\right].
\end{eqnarray}

Since the region of integration is $\mathcal{M}$, we can derive the integral over the singular point, but the Hamiltonian integral does not cover the whole region $\mathcal{M}$. Since the Hamiltonian density is linear and does not depend on $t$, we can transform
\begin{eqnarray}
\int_{\mathcal{M}}\mathcal{H}d^3x\delta_\Sigma = \int_0^\beta \left[\int \mathcal{H} \hspace{1mm} d^2x\right] \hspace{1mm} dt\hspace{1mm}\delta_\Sigma = \beta \int \mathcal{H} \hspace{1mm} d^2x\hspace{1mm} \delta_\Sigma = \int_\Sigma \mathcal{H} d\Sigma.
\end{eqnarray}

We get
\begin{eqnarray}
\beta \int \mathcal{H}_{\mathcal{M}_{\text{singular}}}\hspace{1mm}d^2x \hspace{1mm}\delta_\Sigma &=& \int_\Sigma \mathcal{H}_{\mathcal{M}_{\text{singular}}} d\Sigma, \no\\
\beta  H_{\mathcal{M}_{\text{singular}}}&=& \int_\Sigma \mathcal{H}_{\mathcal{M}_{\text{singular}}} d\Sigma,\no\\
H_{\mathcal{M}_{\text{singular}}}&=&\frac{1}{\beta} \int_\Sigma \mathcal{H}_{\mathcal{M}_{\text{singular}}} d\Sigma.
\end{eqnarray}

Since
\begin{eqnarray}
H_{\mathcal{M}_{\text{singular}}}&=&
-\frac{(1-\alpha)}{4 G\beta}\int_\Sigma d\Sigma\sqrt{-g}\left\{
1 +\frac{3}{4m^2}R
-\frac{1}{m^2}
R_{\mu\nu}\left(n^\mu n^\nu \right)\right\}\no\\
&&-\frac{\pi(1-\alpha)^2}{4 G\beta}\int_\Sigma d\Sigma\sqrt{-g}\delta_\Sigma
\left\{\hspace{2mm}
\frac{3}{2m^2}
-\frac{1}{m^2}(n^\mu n^\nu)(n_\mu n_\nu)
\right\}.\label{40}
\end{eqnarray}

The second term in Eq.\ref{40} does not contribute to the energy so we can eliminate it. We examine the surface $\Sigma$ in which the spherical cone singularity from the surface $[t,r]$ is at $r_H$, since the singularity appears in the time component, the time parameter $t$ does not contribute, and since the singularity appears at the horizon is definite, the space parameter $r$ does not change, so it does not contribute, only the parameter $\phi$ contributes to the integral over the surface $\Sigma$, with the orthogonal coordinates describing the tangent at the singularity being
\begin{eqnarray}
n_1^\mu = (\sqrt{g^{00}},0,0),\hspace{10mm}
n_2^\mu = (0,\sqrt{g^{11}},0).
\end{eqnarray}

The integral of the singularity is
\begin{eqnarray}
H_{\text{singular}}&=&
-\frac{(1-\alpha)}{4 G\beta}\int_0^{2\pi} d\phi\sqrt{g_{22}|_{r=r_H}}\left[
1 +\frac{3}{4m^2}R
-\frac{1}{m^2}
R_{\mu\nu}\left(n^\mu n^\nu \right)\right]\no\\
&=&
-\frac{\pi(1-\alpha)}{2 G\beta}r_H\left(
1 -\frac{1}{2m^2l^2}\right).\label{42}
\end{eqnarray}

With the thermodynamic investigation of the phase transition, it is reasonable that the singularity deficit angle can be related to the Hawking temperature. Since the singularity is in the time component of the metric, it can be related to the Hawking time. We choose to fix the coefficient $\alpha=1$ or $\alpha=\dis\beta \frac{\sqrt{M}}{2\pi l}$. Therefore, Eq.\ref{42} is rewritten as follows
\begin{eqnarray}
H_{\text{singular}}&=&\left(\frac{M}{4G}-\frac{\pi\l\sqrt{M}}{2G\beta}\right)\left(
1 -\frac{1}{2m^2l^2}\right).
\end{eqnarray}

We get the free energy of the black hole state, namely
\begin{eqnarray}
F_{bh}=H_{bh}&=&H_{bulk}+H_{singular}\no\\
&=&\left(\frac{M}{8G}-\frac{\pi\l\sqrt{M}}{2G\beta}\right)\left(
1 -\frac{1}{2m^2l^2}\right).
\end{eqnarray}

We examine $M<0$, because of the general structure of the thermal soliton, we consider the surface $\Sigma$ in which the conical singularity at $r=0$ is composed of the surface $[r,\phi]$, therefore the space represented by the metric is no longer a curved space. With $r=0$ the rotational component $\phi$ no longer contributes, it follows that $r$ is definite hence the spatial parameter $r$ does not change and does not contribute, thus only the time parameter $t$ contributes to the integral on the surface $\Sigma$, with the orthogonal coordinates describing the tangent at the singularity being
\begin{eqnarray}
n_1^\mu = (0,\sqrt{g^{11}},0),\hspace{10mm}
n_2^\mu = (0,0,\sqrt{g^{22}}).
\end{eqnarray}

The integral of the singularity has the form
\begin{eqnarray}
H_{\text{singular}}^{soliton}&=&
-\frac{(1-\alpha)}{4 G\beta}\int_0^{\beta} dt\sqrt{g_{00}|_{r=0}}\left[
1 +\frac{3}{4m^2}R
-\frac{1}{m^2}
R_{\mu\nu}\left(n^\mu n^\nu \right)\right]\no\\
&=&
-\frac{(1-\alpha)}{4 G}\sqrt{-M}\left(
1 -\frac{1}{2m^2l^2}\right).
\end{eqnarray}

With the phase transition investigated at $M<0$, it is reasonable that the deficit angle will be related to $M$, we choose $\alpha=\sqrt{-M}$, so that
\begin{eqnarray}
H_{\text{singular}}^{soliton}&=&-\frac{(1-\sqrt{-M})}{4 G}\sqrt{-M}\left(
1 -\frac{1}{2m^2l^2}\right).
\end{eqnarray}

The thermal soliton state free energy is given by
\begin{eqnarray}
F_{soliton}=H_{soliton}&=&H_{bulk}+H_{\text{singular}}^{soliton}\no\\
&=&-\frac{1}{8G}\left(
M+2\sqrt{-M}
\right)
\left(1-\frac{1}{2m^2l^2}\right).
\end{eqnarray}

\begin{figure}[!htb]
\centering
\includegraphics[width=\textwidth]{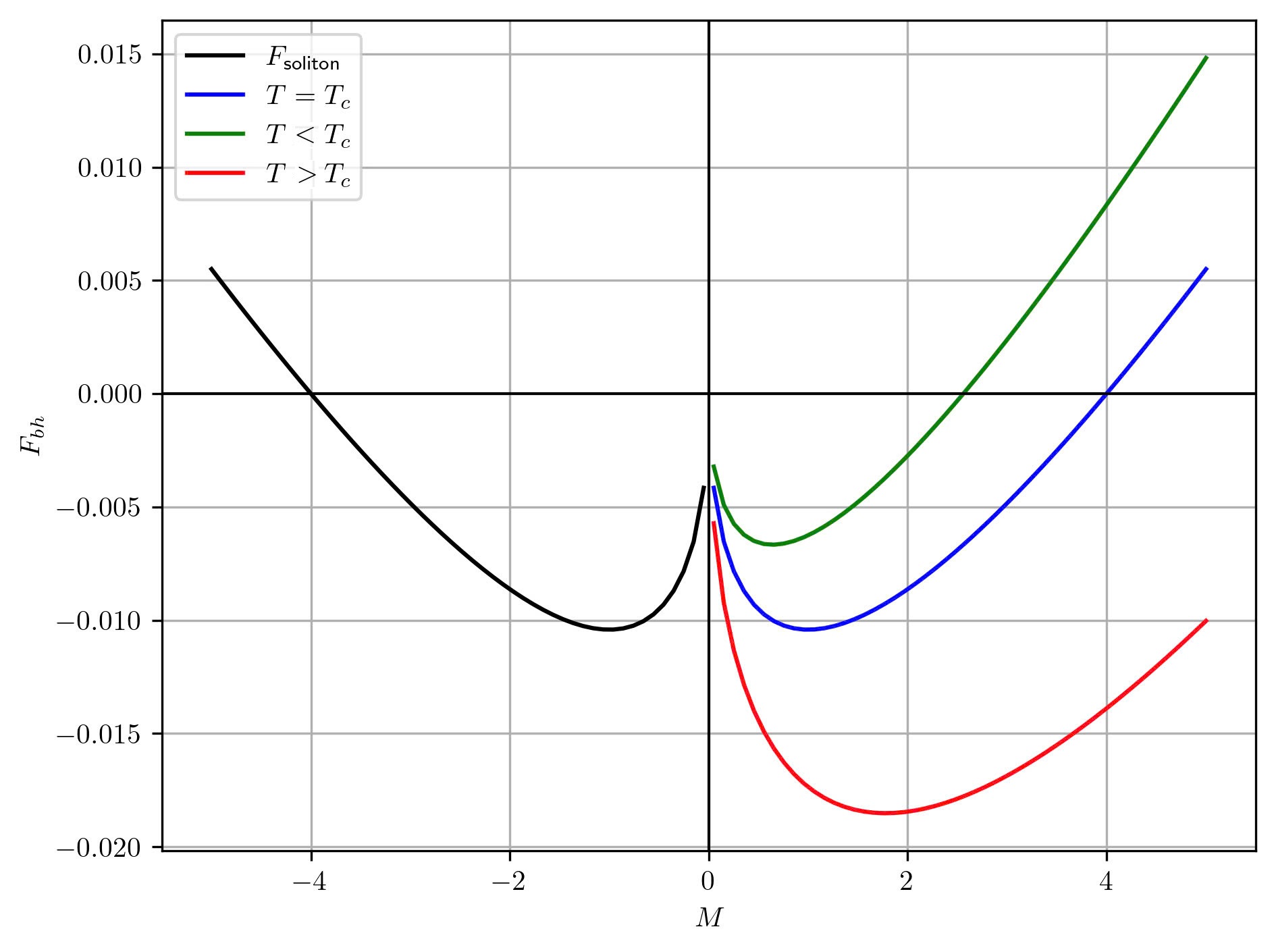}
\caption{The off-shell free energy of the soliton and the BTZ black hole at a fixed temperature $T$ with respect to $M$.}\label{2}
\end{figure}

The two ends of $F_{bh}$ and $F_{soliton}$ shown in Fig.\ref{2} are joined at $M=0$ representing a phase transition, and since the thermal soliton state line is a continuous function connected to the black hole state line, thus the phase transition in this case is continuous.

\subsection{Evaluation}

Consistent with the result obtained directly from the Euclidean action, the on-shell Hamiltonian in Hamiltonian formalism yields the critical temperature $T_C = \dis \frac{1}{2\pi l}$, at which the transition between the thermal soliton state and the black hole state occurs. At the critical temperature, the thermodynamic dominance switches between the two phases, indicating the first-order nature of the Hawking-Page phase transition in the on-shell configuration.

For the off-shell configuration, the black hole geometry is characterized by a conical singularity with a deficit angle, while the soliton phase corresponds to $M<0$. Unlike the on-shell case, the off-shell configuration forms a continuous connection between the soliton and black hole phases through $M=0$. 

Therefore, our results demonstrate that the Hamiltonian formalism can provide an alternative approach to studying the Hawking-Page phase transition, with the on-shell results reproducing the corresponding phase structure obtained from the conventional Euclidean action method for the BTZ black hole. This agreement motivates extending the Hamiltonian analysis to more general black-hole backgrounds. We then extend our investigation to the on-shell Hamiltonians of the non-rotating charged Reissner-Nordstr\"{o}m black hole and further analyze its off-shell thermodynamic phase structure.

\section{Reissner-Nordstr\"{o}m black hole}\label{sec4}

The Reissner-Nordstr\"{o}m black hole is a spherically symmetric, non-rotating, electrically charged black hole solution in four-dimensional spacetime black hole, where the spacetime coordinates are $x^\mu = (t,r,\theta,\phi)$, with the spacetime index $\mu = 0,1,2,3$. The metric of the Reissner-Nordstr\"{o}m black hole in the AdS has the form \cite{RN3}
\begin{eqnarray}
ds^2_{RN}=-f(r)dt^2 + \frac{1}{f(r)}dr^2+r^2d\theta^2 + r^2\sin^2\theta d\phi^2,
\end{eqnarray}
where $f(r)=1-\dis\frac{2M}{r}+\frac{Q^2}{r^2}+\frac{r^2}{l^2}$, with the ADM mass $M$, the AdS radius $l$ and the electric charge $Q$. The event horizon, $r_H$, is the solution of the equation
\begin{eqnarray}
r_H^4+l^2r_H^2-2Ml^2r_H+Q^2l^2=0,
\end{eqnarray}

which takes a complicated form and requires $r_H$ to be equal to or greater than a certain positive constant. Since $r_H=r_H(M)$, it follows that $M$ must be equal to or greater than a corresponding positive minimum threshold.

Along with that, we have the Hawking temperature, \begin{equation}
T_H=\dis\frac{1}{4\pi}\left.\dis\frac{df}{dr}\right|_{r_H}
=\frac{1}{2\pi l^2 r_H^3}\left(Ml^2r_H-Q^2l^2+r_H^4\right).
\end{equation}

We use the Einstein-Maxwell theory to investigate the effects of electric charge on the Hawking-Page phase transition. Constructing the Lagrangian density, the Einstein-Hilbert and Maxwell actions are given by
\begin{eqnarray}
I_{bulk}=I_{EH}&=&\frac{1}{16\pi }\int_\mathcal{M}d^4x\sqrt{-g}(R-F_{\mu\nu}F^{\mu\nu}),
\end{eqnarray}
where $F_{\mu\nu}$ is the electromagnetic field tensor with the vector field $A_\mu$,
\begin{eqnarray}
F_{\mu\nu}=\partial_\mu A_\nu - \partial_\nu A_\mu, \hspace{10mm}A=A_\mu dx^\mu = -\frac{Q}{r}dt.
\end{eqnarray}

To get the full Euclidean action, we need to add the term $I_{GGH}$ which is the surface action of the hypersurface $\partial \mathcal{M}$ with boundary conditions, $r \rightarrow \infty$, after integration. To normalize the action, we need to add the term $I_{ct}$ to eliminate divergences. We have the total action:
\begin{eqnarray}
I&=&I_{\text{bulk}}+I_{GGH}+I_{ct},\\
I_{GGH}&=&-\frac{1}{8\pi}\int_{\partial \mathcal{M}}d^3x\sqrt{-\gamma}K,\\
I_{ct}&=&\frac{1}{8\pi}\int_{\partial \mathcal{M}}d^3x\sqrt{-\gamma}K_0.
\end{eqnarray}

From the above actions we can decompose the Lagrangian of the system,
\begin{eqnarray}
I=\int_\mathcal{M}d^4x\hspace{1mm} \mathcal{L}_\mathcal{M}+ \int_{\partial \mathcal{M}}d^3x \hspace{1mm}\mathcal{L}_{\partial \mathcal{M}}.
\end{eqnarray}
with 
\begin{eqnarray}
\mathcal{L}_\mathcal{M} &=&\frac{1}{16\pi}\sqrt{-g}\left(R-F_{\mu\nu}F^{\mu\nu}\right),\\
\mathcal{L}_{\partial \mathcal{M}}&=&-\frac{1}{8\pi}\sqrt{-\gamma}(K-K_0).
\end{eqnarray}

Similar to the BTZ black hole, the Hamiltonian of the RN black hole is then decomposed into the bulk and boundary contributions. The metric is stationary and contains no $dtdx^a$ cross terms, hence $p^{rs}\partial_0 g_{rs} = 0$
\begin{eqnarray}
H_{\mathcal{M}}&=\dis\int_{\mathcal{M}} d^3x\hspace{1mm} (p^{rs}_{_{_{\mathcal{M}}}}\partial_0 g_{rs_{\mathcal{M}}}-\mathcal{L}_\mathcal{M})=&-\int_{\mathcal{M}} d^3x\hspace{1mm} \mathcal{L}_\mathcal{M},\\
H_{\partial \mathcal{M}}&=\dis\int_{\mathcal{M}} d^2x\hspace{1mm} (p^{rs}_{_{_{\partial \mathcal{M}}}}\partial_0 g_{rs_{\partial \mathcal{M}}}-\mathcal{L}_{\partial \mathcal{M}})=&-\int_{\partial\mathcal{M}} d^2x\hspace{1mm} \mathcal{L}_{\partial\mathcal{M}}.
\end{eqnarray}

\subsection{The Hamiltonian of the black hole}

The metric derived from $ds^2_{RN}$ in the space $\mathcal{M}$ has the form
\begin{eqnarray}
g_{\mu\nu}=\begin{pmatrix}
-f(r) 	&0 					&0		&0\\
0		&\frac{1}{f(r)}		&0		&0\\
0		&0					&r^2	&0\\
0		&0					&0		&r^2\sin^2\theta
\end{pmatrix},\hspace{2mm}g^{\mu\nu}=\begin{pmatrix}
-\frac{1}{f(r)} 	&0 			&0						&0\\
0					&f(r)		&0						&0\\
0					&0			&\frac{1}{r^2}			&0\\
0					&0			&0						&\frac{1}{r^2\sin^2\theta}
\end{pmatrix}.
\end{eqnarray}

The Ricci tensor is calculated in the matrix form,
\begin{eqnarray}
R_{\mu\nu}=\left(\dis\frac{Q^2}{r^4}-\frac{3}{l^2}\right)\begin{pmatrix}
-g_{00}	&0			&0			&0\\
0		&-g_{11}	&0			&0\\
0		&0			&g_{22}		&0\\
0		&0			&0			&g_{33}
\end{pmatrix}.
\end{eqnarray}

And the contribution of the electromagnetic field $F_{\mu\nu}F^{\mu\nu}=\dis \frac{Q^2}{r^4}$. The gravitational Lagrangian on the $\mathcal{M}$ surface is
\begin{eqnarray}
\mathcal{L}_\mathcal{M}&=&-\frac{1}{8\pi}
\frac{Q^2}{r^2}\sin\theta.
\end{eqnarray}

The metric is decomposed with respect to the boundary hypersurface $\partial\mathcal{M}$, defined by $r=const$, with the spacetime coordinates $x^\mu=(t,r,\theta,\phi)$, where $t$, $\theta$, and $\phi$ are the intrinsic coordinates on $\partial\mathcal{M}$, labeled by $i,j$. The metric can then be written as
\begin{eqnarray}
ds^2_{RN}&=&N^2dr^2 + \gamma_{ij}(dx^i + N^idr)(dx^j + N^jdr),
\end{eqnarray}
where the shift vector vanishes, $N^i = 0$, due to the absence of cross terms of the form $dx^i dr$, with the lapse function $N$, and the induced metric $\gamma_{ij}$ are
\begin{eqnarray}
N=\sqrt{\frac{1}{f(r)}},\hspace{10mm}\gamma_{ij}&=&\begin{pmatrix}
-f(r)	&0		&0\\
0	&r^2	&0\\
0	&0		&r^2\sin^2\theta
\end{pmatrix},
\end{eqnarray}

and the extrinsic curvature, $K_{ij}=-\dis\frac{\sqrt{f(r)}}{2}\partial_r \gamma_{ij}$, and the normalization term $K_0 =\dis\frac{\sqrt{f}}{f}\frac{r}{l^2}- \frac{\sqrt{f}}{r^2}\partial_r r^2$. The Lagrangian on the boundary $\partial \mathcal{M}$ has the form
\begin{eqnarray}
\mathcal{L}_{\partial \mathcal{M}}&=&\frac{1}{8\pi}\left(M-\frac{Q^2}{r}\right)\sin\theta.
\end{eqnarray}

The Hamiltonian of the RN black hole takes the form
\begin{eqnarray}
	H_{bh}&=&H_{\mathcal{M}}+H_{\partial \mathcal{M}}\no\\
	&=&-\frac{1}{2}\left(M-\frac{Q^2}{r_H}\right),
\end{eqnarray}
where
\begin{eqnarray}
H_{\mathcal{M}}&=&-\frac{Q^2}{2}\left(
\frac{1}{r}-\frac{1}{r_H}
\right),\\
H_{\partial\mathcal{M}}&=&-\frac{1}{2}\left(M-\frac{Q^2}{r}\right).
\end{eqnarray}

The Helmholtz free energy obtained directly from the Euclidean action is given by
\begin{eqnarray}
F_{bh}&=&-\frac{1}{2}\left(M-\frac{Q^2}{r_H}\right),
\end{eqnarray}

The agreement between the Hamiltonian obtained from the Hamiltonian formalism and the Helmholtz free energy obtained independently from the Euclidean action further supports our previous conclusion that the Hamiltonian formalism provides an alternative approach to studying Hawking-Page transitions. In particular, this result demonstrates that the formalism can also be applied to charged black holes, such as the RN black hole.

\subsection{On-Shell}

In the on-shell configuration, we consider the thermal soliton state by setting $M=0$ and $Q=0$, for which the metric reduces to the AdS form. The Hamiltonian of the soliton state is then
\begin{eqnarray}
F_{soliton}=H_{soliton}=H_{\mathcal{M}_{soliton}} + H_{\partial \mathcal{M}_{soliton}}=0.
\end{eqnarray}

The phase transition occurs when $F_{bh}=F_{soliton}$, therefore $M=\dis\frac{Q^2}{r_H}$. We get the critical temperature $T_C=\dis\frac{r_H}{2\pi l^2}$.
\begin{figure}[!htb]
\centering
\includegraphics[width=\textwidth]{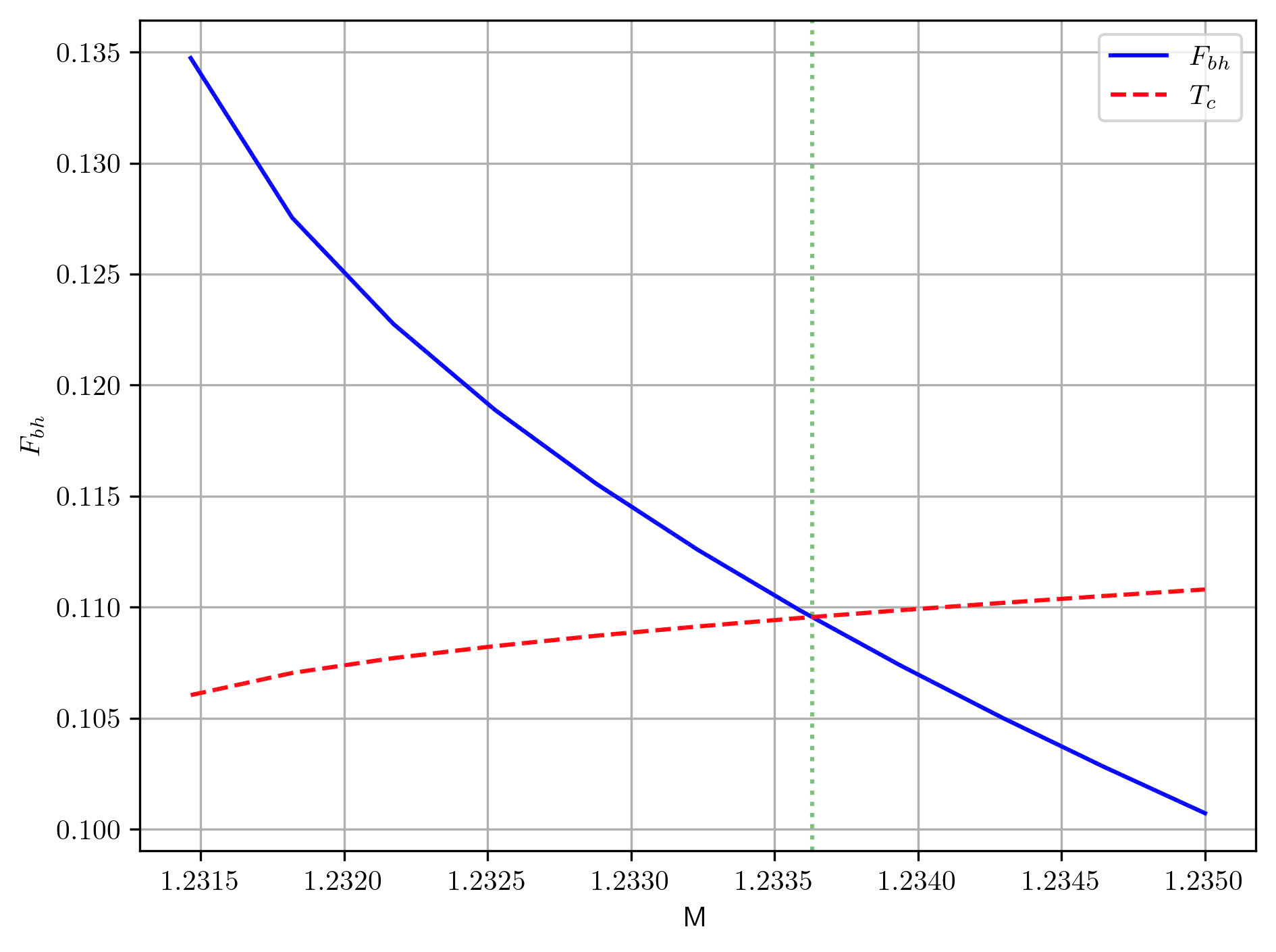}
\caption{The on-shell free energy and critical temperature at a fixed charge of the RN black hole as a function of $M$.}\label{3}
\end{figure}

Fig.\ref{3} indicates that the on-shell free energy has a lower bound representing the minimum mass threshold, which exists due to the presence of the electric charge, $Q$. Moreover, the phase transition occurs at the intersection of $F_{bh}$ line and $T_C$ line, from the thermal soliton state to the black hole state. Similar to the BTZ black hole, the phase transition in on-shell spacetime occurs instantaneously and therefore the Hawking-Page phase transition is a first-order phase transition. 

\subsection{Off-Shell}

We consider the off-shell configuration, represented by the conical geometry $\mathcal{M}_\alpha$ with deficit angle $2\pi(1-\alpha)$ and the singularity set $\Sigma$, to examine its thermodynamic properties away from equilibrium. Thus
\begin{eqnarray}
\mathcal{L}_\mathcal{M} &=&\frac{1}{16\pi}\sqrt{-g}\left[
R^{(\alpha)}-F_{\mu\nu}F^{\mu\nu}
\right]\no\\
&=&\frac{1}{16\pi}\sqrt{-g}\left[
4\pi(1-\alpha)\delta_\Sigma - \frac{Q^2}{r^4}
\right]\no\\
&=&\mathcal{L}_{\mathcal{M}_{\text{bulk}}}+\mathcal{L}_{\mathcal{M}_{\text{singular}}}.
\end{eqnarray}

Considering only $H_{\mathcal{M}_{\text{singular}}}$, we separate the terms according to $\delta_\Sigma$,
\begin{eqnarray}
H_{\mathcal{M}_{\text{singular}}}&=&-\frac{(1-\alpha)}{4\beta}\int_\Sigma d\Sigma\sqrt{|g|}.
\end{eqnarray}

We examine the singular set $\Sigma$ in which the cone at $r=r_H$ is made up of the surface $[t,r]$ rotating in $\Sigma$. At this point, the space parameter is fixed and therefore does not contribute to the integral. Similarly the element before the time parameter also disappears, so time no longer contributes to the integral. Thus, only the zenith angle $\theta$ and the azimuth angle $\phi$ contribute, with the coordinates describing the tangent being
\begin{eqnarray}
n_1^\mu = (\sqrt{g^{00}},0,0,0),\hspace{10mm}
n_2^\mu = (0,\sqrt{g^{11}},0,0).
\end{eqnarray}

We choose to fix the coefficient $\alpha=1$ or $\alpha=\dis T\beta= \frac{1}{2\pi l^2 r_H^3}\left(Ml^2r_H-Q^2l^2+r_H^4\right)\beta$,

\begin{eqnarray}
H_{\text{singular}}&=&-\frac{(1-\alpha)}{4\beta}r^2_H\int_0^\pi \sin\theta d\theta \int_0^{2\pi}d\phi\no\\
&=&-\frac{\pi}{\beta}r_H^2 
+\frac{r_H^3}{2l^2}
+\frac{1}{2}\left(M-\frac{Q^2}{r_H}\right).
\end{eqnarray}

The free energy of the black hole state is
\begin{eqnarray}
F_{bh}=H_{bh}&=&H_{bulk}+H_{singular}\no\\
&=&\frac{r_H^3}{2l^2}-\frac{\pi}{\beta}r_H^2.
\end{eqnarray}

We consider $M$ such that $T<T_C$. Because of the general structure of the thermal soliton, we consider the surface $\Sigma$ in which the conical singularity at $r=\dis\frac{Q^2}{2M}$ is composed of tangent $r$. At this point, the space represented by the metric is no longer a curved space, because $r=\dis\frac{Q^2}{2M}$ is defined so the spatial parameter $r$ does not change so it does not contribute. Consequently, the parameters that still contribute to the integral on the surface $\Sigma$ include: time $t$, zenith angle $\theta$, azimuth angle $\phi$, with the coordinates describing the tangent at the singularity being
\begin{eqnarray}
n_1^\mu = (0,\sqrt{g^{11}},0,0).
\end{eqnarray}

The integral of the singularity has the form
\begin{eqnarray}
H_{\text{singular}}^{soliton}&=&
-\left.\frac{(1-\alpha)}{4 \beta}r^2\sqrt{f}\right|_{r=\frac{Q^2}{2M}}\int_0^{\beta} dt \int_0^\pi \sin\theta d\theta \int_0^{2\pi}d\phi\no\\
&=&-\pi(1-\alpha)\frac{Q^4}{4M^2}\sqrt{1+\frac{Q^4}{4M^2l^2}}.
\end{eqnarray}

Considering the phase transition at $T<T_C$, it is reasonable that the deficit angle will be related to $M$, and since there exists a minimum mass threshold we choose $\alpha=\sqrt{2\pi-M}$.

\begin{eqnarray}
H_{\text{singular}}^{soliton}&=&-\pi(1-\sqrt{2\pi-M})\frac{Q^4}{4M^2}\sqrt{1+\frac{Q^4}{4M^2l^2}}.
\end{eqnarray}

The thermal soliton state free energy is given by
\begin{eqnarray}
H_{soliton}=F_{soliton}&=&H_{bulk}+H_{\text{singular}}^{soliton}\no\\
&=&-\frac{1}{2}\left(M-\frac{Q^2}{r_H}\right)-\pi(1-\sqrt{2\pi-M})\frac{Q^4}{4M^2}\sqrt{1+\frac{Q^4}{4M^2l^2}}.
\end{eqnarray}

\begin{figure}[!htb]
\centering
\includegraphics[width=\textwidth]{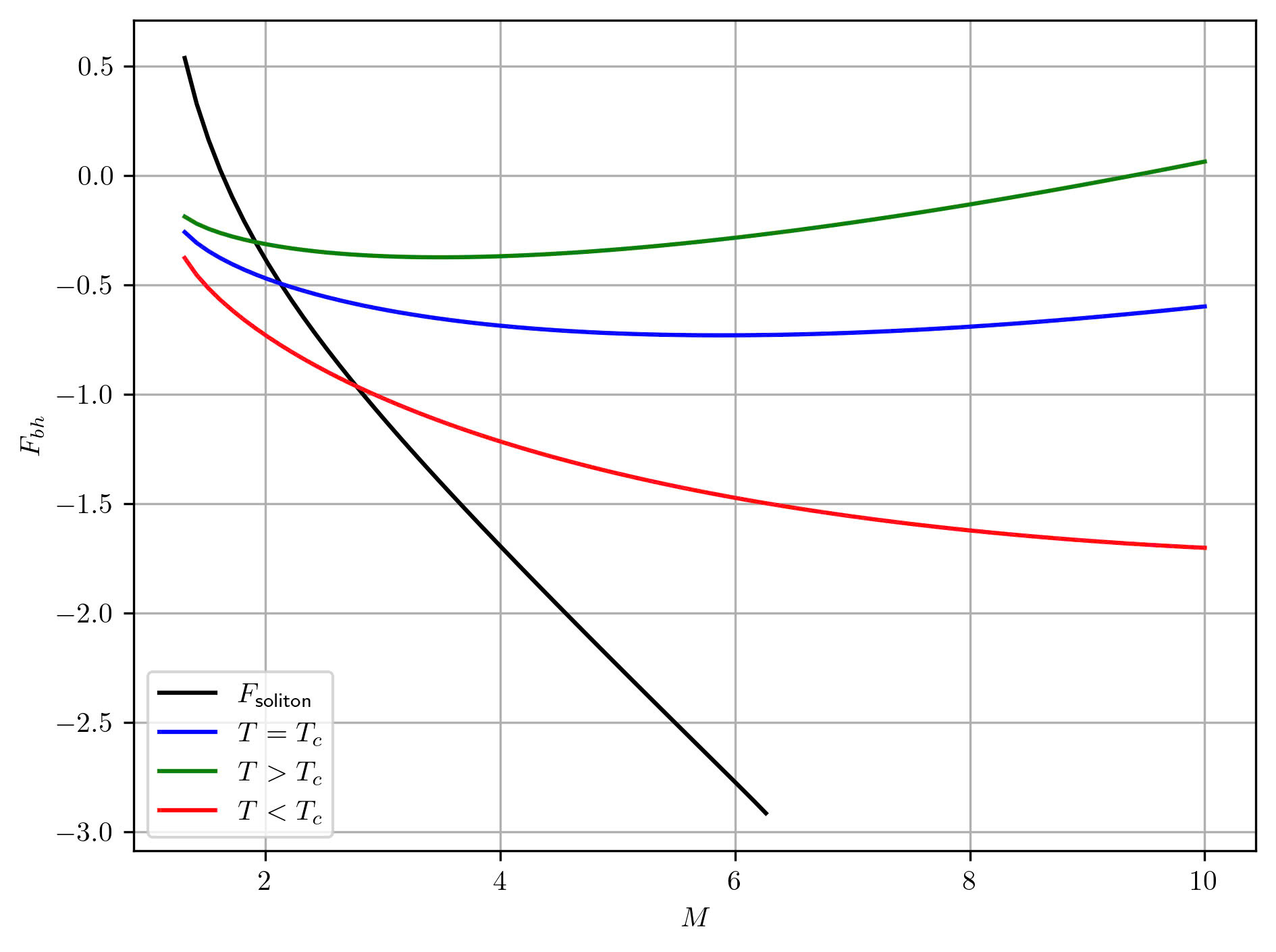}
\caption{The off-shell free energy of the soliton and the RN black hole at a fixed temperature $T$ and electric charge $Q$ with respect to $M$.}\label{4}
\end{figure}

From Fig.\ref{4}, the free energy in the off-shell case has a lower mass bound for both the thermal soliton state and the black hole state which represents the mass threshold. This threshold exists due to the presence of electric charge, $Q$, similar to the on-shell case. Moreover, since the free energy of the thermal soliton state has an upper mass bound, arising from the minimum mass threshold, it follows that the thermal soliton state is confined to a certain range. Notably, the free energies of the two states intersect, indicating the relevance of both states within the same thermodynamic ensemble. Thus, the phase transition involves a redistribution of the thermodynamic dominance between the thermal soliton and the black hole states, rather than a clear-cut change from one state to the other, allowing it to be regarded as a continuous phase transition.

\subsection{Evaluation}

In the on-shell configuration, we obtain the critical temperature, $T_C = \dis \frac{r_H}{2\pi l}$, at which the transition between the black hole and the thermal soliton phases occurs. Importantly, the presence of electric charge $Q$ imposes a minimum mass threshold for the existence of the event horizon. Consequently, the transition occurs abruptly at the critical temperature, indicating a first-order phase transition. This result is consistent with the previously established result obtained directly from the Euclidean action, further demonstrating that the Hamiltonian formalism provides an alternative framework for studying Hawking-Page phase transitions.

Our off-shell analysis reveals a continuous transition between the thermal soliton and the black hole phases. The electric charge imposes a minimum mass threshold on the system, restricting the range in which the thermal soliton phase is relevant and allowing both the thermal soliton and the black hole states to be considered within the same thermodynamic ensemble. As the temperature increases, the black hole state progressively becomes thermodynamically dominant, whereas the thermal soliton state becomes dominant as the temperature decreases, within its relevant range. The resulting continuous behavior is consistent with a second-order phase transition.

Thus, our on-shell results demonstrate that the Hamiltonian formalism can be applied to studying the Hawking-Page phase transition in charged black holes. In addition, our off-shell analysis provides new insights into the phase transition, in which the quantum nature of the system is reflected in the shifting dominance of both the thermal soliton and black hole states within the same ensemble. Building on these results, we further apply the Hamiltonian formalism to investigate the on-shell Hawking-Page phase transition of the Kerr-Newman black hole and further analyze its off-shell thermodynamic phase structure.

\section{Kerr-Newman black hole}\label{sec5}

The Kerr-Newman black hole \cite{KN1} is a stationary, axisymmetric, rotating, charged black hole solution in four-dimensional spacetime, where the spacetime coordinates are $x^\mu = (t,r,\theta,\phi)$, with the spacetime index $\mu = 0,1,2,3$. In the AdS spacetime using the Boyer-Lindquist coordinates, the metric has the form
\begin{eqnarray}
ds^2_{KN}&=&\hspace{3mm}
\left(-\frac{\Delta_r}{\rho^2}+\frac{a^2\Delta_\theta \sin^2\theta}{\rho^2}\right)dt^2
+\left[\frac{a\Delta_r\sin^2\theta}{\Xi\rho^2}-\frac{a(r^2+a^2)\Delta_\theta \sin^2\theta}{\Xi\rho^2}\right]dtd\phi\no\\
&&+\frac{\rho^2}{\Delta_r}dr^2
+\frac{\rho^2}{\Delta_\theta}d\theta^2
+\left[\frac{a\Delta_r\sin^2\theta}{\Xi\rho^2}-\frac{a(r^2+a^2)\Delta_\theta \sin^2\theta}{\Xi\rho^2}\right]d\phi dt\no\\
&&+\left(-\frac{a^2\Delta_r \sin^4\theta}{\Xi^2\rho^2}+\frac{(r^2+a^2)^2\Delta_\theta \sin^2\theta}{\Xi^2\rho^2}\right)d\phi^2,
\end{eqnarray}
where
\begin{eqnarray}
\begin{matrix}
\Delta_r =&\left(r^2+a^2\right)\left(1+\dis\frac{r^2}{l^2}\right)-2Mr +\left(q_e^2+q_m^2\right),\\
\Delta_\theta =&1-\dis\frac{a^2}{l^2}\cos^2\theta,
\end{matrix}\hspace{10mm}
\begin{matrix}
\rho^2 =&r^2+a^2\cos^2\theta,\\
\Xi =&1-\dis\frac{a^2}{l^2}.
\end{matrix}
\end{eqnarray}

where $M$ is the ADM mass, $l$ is the AdS radius, $q_e$ and $q_m$ are electromagnetic charges, and $a$ is the rotational parameter. The event horizon, $r_H$, is the solution of the equation below:
\begin{eqnarray}
\left(r_H^2+a^2\right)\left(1+\dis\frac{r_H^2}{l^2}\right)-2Mr_H +\left(q_e^2+q_m^2\right)=0,
\end{eqnarray}

Similar to the RN black hole, $r_H$ is complicated and imposes the minimum mass for the event horizon to exist due to the presence of charge and rotation.

The Hawking temperature ($T_H$) has the form
\begin{eqnarray}
T_H=\frac{1}{4\pi}\frac{r_H}{\left(r_H^2+a^2\right)^2}\left[1+\frac{a^2}{l^2}+3\frac{r_H^2}{l^2}-\frac{a^2-\left(q_e^2+q_m^2\right)}{r_H^2}\right].
\end{eqnarray}

Taking AdS space as the reference background, the mass $m$ and angular momentum $J$ can be defined by means of Komar integrals, using the Killing vectors $\partial_t / \Xi$ and $\partial_\phi^4$,
\begin{eqnarray}
m = \frac{M}{\Xi}, \hspace{10mm}J = \frac{aM}{\Xi}.
\end{eqnarray}

We use the Einstein-Maxwell theory to investigate the effects of charge and rotation on the Hawking-Page phase transition. We examine the Einstein-Hilbert and Maxwell actions to construct the Lagrangian density
\begin{eqnarray}
I_{bulk}=I_{EH}&=&\frac{1}{16\pi }\int_\mathcal{M}d^4x\sqrt{-g}(R-2\Lambda -F_{\mu\nu}F^{\mu\nu}),
\end{eqnarray}

The vector field $A_\mu$ is written in terms of the vierbein basis as a convenient orthonormal basis
\begin{eqnarray}
A=-\frac{q_e r}{\rho \sqrt{\Delta_r}}e^0
-\frac{q_m \cos\theta}{\rho \sqrt{\Delta_\theta}\sin\theta}e^3,
\end{eqnarray}

its exterior derivative leads to the expression for $F_{\mu\nu}$
\begin{eqnarray}
F&=&-\frac{1}{\rho^4}\left[
q_e\left(r^2-a^2\cos^2\theta\right)+2q_m r a \cos\theta
\right]e^0 \wedge e^1\no\\
&&+\frac{1}{\rho^4}\left[
q_m\left(r^2-a^2\cos^2\theta\right)-2q_e r a \cos\theta
\right]e^2 \wedge e^3,
\end{eqnarray}
with vierbein vectors,
\begin{eqnarray}
\begin{matrix}
e^0=&\dis\frac{\sqrt{\Delta_r}}{\rho}\left(
dt-\frac{a\sin^2\theta}{\Xi}d\phi
\right),\\
e^3=&\dis\frac{\sqrt{\Delta_\theta}\sin\theta}{\rho}\left(
adt-\frac{r^2+a^2}{\Xi}d\phi
\right),
\end{matrix}\hspace{10mm}
\begin{matrix}
e^1=&\dis\frac{\rho}{\sqrt{\Delta_r}},\\
e^2=&\dis\frac{\rho}{\sqrt{\Delta_\theta}}.
\end{matrix}
\end{eqnarray}

The total action is
\begin{eqnarray}
I&=&I_{\text{bulk}}+I_{GGH}+I_{ct},\\
I_{GGH}&=&-\frac{1}{8\pi }\int_{\partial \mathcal{M}}d^3x\sqrt{-\gamma}K,\\
I_{ct}&=&\frac{1}{8\pi }\int_{\partial \mathcal{M}}d^3x\sqrt{-\gamma}K_0.
\end{eqnarray}

Thus, we can decompose the Lagrangian into the bulk and boundary contributions,
\begin{eqnarray}
I=\int_\mathcal{M}d^4x\hspace{1mm} \mathcal{L}_\mathcal{M}+ \int_{\partial \mathcal{M}}d^3x \hspace{1mm}\mathcal{L}_{\partial \mathcal{M}},
\end{eqnarray}
where
\begin{eqnarray}
\mathcal{L}_\mathcal{M} &=&\frac{1}{16\pi }\sqrt{-g}\left(R-2\Lambda - F_{\mu\nu}F^{\mu\nu}\right),\\
\mathcal{L}_{\partial \mathcal{M}}&=&-\frac{1}{8\pi }\sqrt{-\gamma}(K-K_0).
\end{eqnarray}

Although the metric is stationary and the canonical momentum $p^{rs}$ is non-zero due to the $g_{03}$ component, the metric components are independent of the time coordinate, so that $\partial_0 g_{\mu\nu} = 0$. Therefore, the term $p^{rs}\partial_0 g_{rs}$ does not contribute to the Hamiltonian.

\begin{eqnarray}
H_{\mathcal{M}}&=\dis\int_{\mathcal{M}}d^3x\hspace{1mm} (p^{rs}_{_{_{\mathcal{M}}}}\partial_0 g_{rs_{\mathcal{M}}}-\mathcal{L}_\mathcal{M})=&-\int_{\mathcal{M}} d^3x\hspace{1mm} \mathcal{L}_\mathcal{M},\\
H_{\partial \mathcal{M}}&=\dis\int_{\mathcal{M}} d^2x\hspace{1mm} (p^{rs}_{_{_{\partial \mathcal{M}}}}\partial_0 g_{rs_{\partial \mathcal{M}}}-\mathcal{L}_{\partial \mathcal{M}})=&-\int_{\partial\mathcal{M}} d^2x\hspace{1mm} \mathcal{L}_{\partial\mathcal{M}}.
\end{eqnarray}

\subsection{The Hamiltonian of the black hole}

The metric derived from $ds^2_{RN}$ in the space $\mathcal{M}$ has the form
\begin{eqnarray}
g_{\mu\nu}=\begin{pmatrix}
g_{00} 		&0 					&0		&g_{03}\\
0			&g_{11}				&0		&0\\
0			&0					&g_{22}	&0\\
g_{30}		&0					&0		&g_{33}
\end{pmatrix},g^{\mu\nu}=\begin{pmatrix}
\dis\frac{g_{33}}{g_{00}g_{33}-g_{03}g_{30}}	&0 			&0						&-\dis\frac{g_{30}}{g_{00}g_{33}-g_{03}g_{30}}\\
0					&\dis\frac{1}{g_{11}}		&0						&0\\
0					&0			&\dis\frac{1}{g_{22}}			&0\\
-\dis\frac{g_{03}}{g_{00}g_{33}-g_{03}g_{30}}					&0			&0						&\dis\frac{g_{00}}{g_{00}g_{33}-g_{03}g_{30}}
\end{pmatrix}.
\end{eqnarray}

With $\Lambda =\dis -\frac{3}{l^2}$, we deduce the Lagrangian on the $\mathcal{M}$ surface,
\begin{eqnarray}
\mathcal{L}_\mathcal{M}&=&-\frac{1}{16\pi }\sqrt{-g}\left(\frac{6}{l^2} + F_{\mu\nu}F^{\mu\nu}\right).
\end{eqnarray}

The same decomposition is applied to the metric of the spacetime manifold $\mathcal{M}$, whose spacetime coordinates are $x^\mu=(t,r,\theta,\phi)$, with respect to the boundary hypersurface $\partial\mathcal{M}$ defined by $r=\mathrm{const}$. Thus, $t$, $\theta$, and $\phi$ are the intrinsic coordinates on $\partial\mathcal{M}$, labeled by $i,j$. We obtain
\begin{eqnarray}
ds^2_{KN}&=&N^2dr^2 + \gamma_{ij}(dx^i + N^idr)(dx^j + N^jdr),
\end{eqnarray}
with
\begin{eqnarray}
N=\sqrt{\frac{\rho^2}{\Delta_r}},\hspace{10mm} N^i = 0,\hspace{10mm}\gamma_{ij}&=&\begin{pmatrix}
\gamma_{00}	&0		&\gamma_{02}\\
0	&\gamma_{11}	&0\\
\gamma_{20}	&0		&\gamma_{22}
\end{pmatrix}.
\end{eqnarray}

And the extrinsic curvature, $K_{ij}=-\dis\frac{\sqrt{f(r)}}{2}\partial_r \gamma_{ij}$, and the normalization term $K_0 =\dis-\frac{\sqrt{\Delta_r}}{2\rho}\left(\gamma^{11}\partial_r \gamma_{11}+\Xi r\right)$. The Lagrangian on the boundary $\partial \mathcal{M}$ has the form
\begin{eqnarray}
\mathcal{L}_{\partial \mathcal{M}}&=&\frac{1}{16\pi \Xi l^2}\left(
2r^3+2a^2r - 2Ml^2\right)\sin\theta.
\end{eqnarray}

We deduce the Hamiltonians,
\begin{eqnarray}
H_{\mathcal{M}}&=&\frac{1}{4\Xi l^2}\left[
2r^3+2a^2r-2r_H^3-2a^2r_H +2\frac{\left(q_e^2-q_m^2\right)r}{a^2+r^2}l^2 - 2\frac{\left(q_e^2-q_m^2\right)r_H}{a^2+r_H^2}l^2
\right],\\
H_{\partial \mathcal{M}}&=&-\frac{1}{4\Xi l^2}\left(
2r^3+2a^2r - 2Ml^2
\right).
\end{eqnarray}

With the boundary conditions for $r \rightarrow \infty$ and transformations based on $\dis\Delta_r (r_H)=\dis\frac{\Delta_r (r_H)}{r_H}=0$, the Hamiltonian of the KN black hole has the form
\begin{eqnarray}
H_{bh}&=&H_{\mathcal{M}}+H_{\partial \mathcal{M}}=F_{bh}\no\\
&=&\frac{1}{4\Xi l^2}\left[
-r_H^3+\Xi l^2 r_H + \frac{a^2l^2}{r_H} + \frac{\left(q_e^2+q_m^2\right)l^2}{r_H}  - 2\frac{\left(q_e^2-q_m^2\right)l^2}{a^2+r_H^2}r_H
\right].
\end{eqnarray}

In agreement with the Helmholtz free energy obtained from the Euclidean approach \cite{KN1}, the Hamiltonian formalism can also be used to investigate the Hawking-Page phase transition of a charged and rotating black hole. This result further supports the applicability of the Hamiltonian formalism to more general black holes.
	
\subsection{On-Shell}

In the on-shell configuration, we consider the thermal soliton with $M=0$, $a=0$ and $q=0$, the metric reduces to the AdS form, which implies $r_H=0$. The soliton state is
\begin{eqnarray}
H_{soliton}=H_{\mathcal{M}_{soliton}} + H_{\partial \mathcal{M}_{soliton}}=0.
\end{eqnarray}

The phase transition occurs when $F_{bh}=F_{soliton}$, which implies $M=\dis\frac{\left(r_H^2+a^2\right)^2+q^2a^2}{r_H^2+a^2}$.

The corresponding critical temperature is then 
\begin{equation}
T_C=\dis\frac{1}{2\pi l^2 }\frac{1}{\left(r_H^2+a^2\right)^2r_H}\left[\frac{r_H^4}{l^2}-a^2+\frac{\left(r_H^2+a^2\right)^2+q^2a^2}{r_H^2+a^2}\right].\label{kntc}
\end{equation}

\begin{figure}[!htb]
\centering
\includegraphics[width=\textwidth]{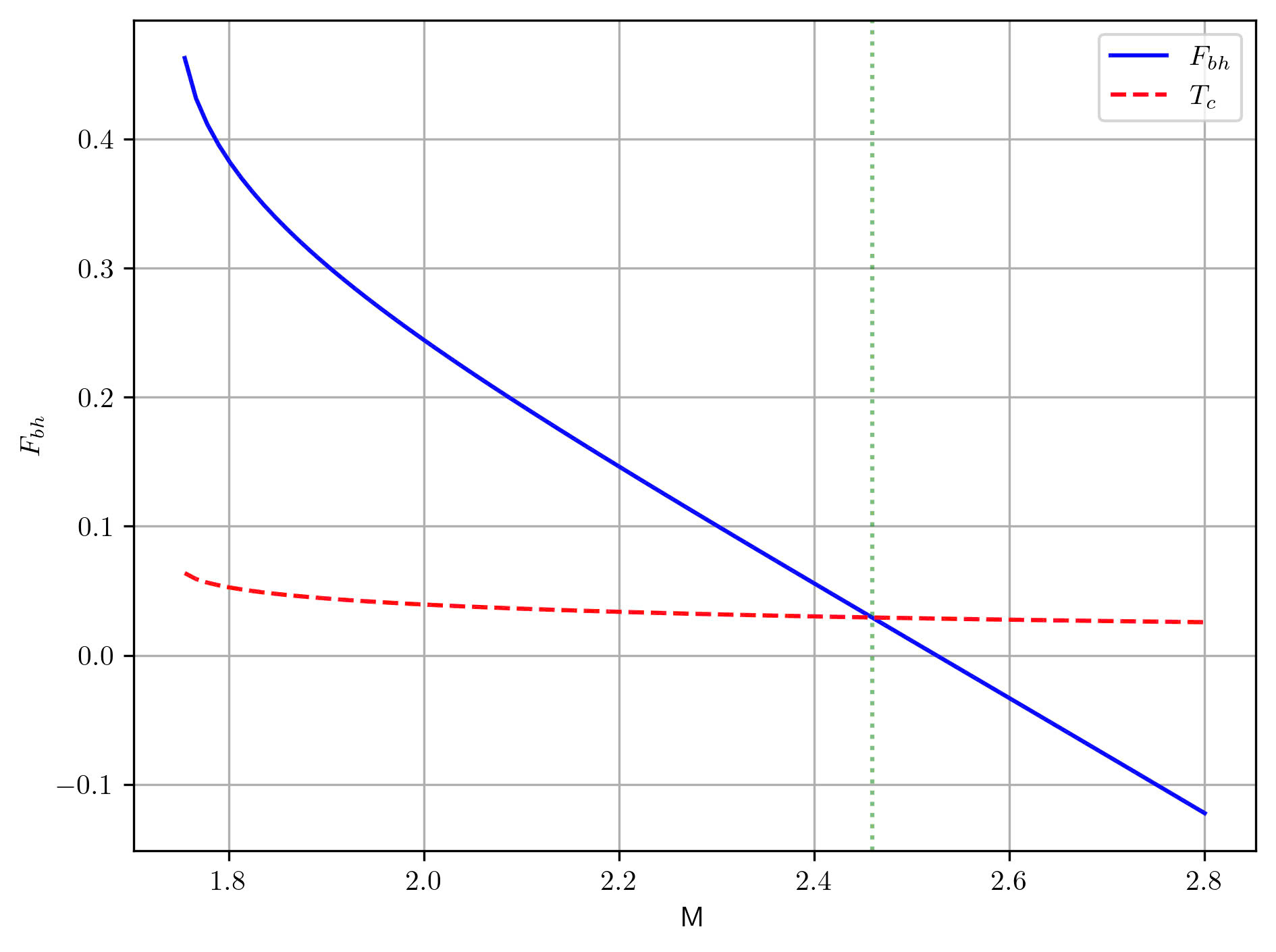}
\caption{The on-shell free energy and critical temperature at a fixed charge and the rotation of the KN black hole, in terms of $M$.}\label{5}
\end{figure}

Firstly, comparing Fig.\ref{5} with Fig.\ref{3}, we observe that the on-shell free energies of the RN black hole and the KN black hole have qualitatively identical forms. Moreover, the presence of charge $q$ and the rotation $a$ set a minimum mass threshold. The phase transition from the thermal soliton state to the black hole state takes place as the $F_{bh}$ line and the $T_C$ line intersect. However, the significance of rotation has not been clearly demonstrated due to the spherical symmetry of the system. The phase transition occurs instantaneously and therefore is a first-order phase transition.

\subsection{Off-Shell}

We consider the off-shell configuration, described by the conical geometry $\mathcal{M}_\alpha$ with deficit angle $2\pi(1-\alpha)$ and the singularity set $\Sigma$, to examine its thermodynamic properties away from equilibrium. Thus
\begin{eqnarray}
\mathcal{L}_\mathcal{M} &=&\frac{1}{16\pi }\sqrt{-g}\left[
R^{(\alpha)}-F_{\mu\nu}F^{\mu\nu}
\right]\no\\
&=&\frac{1}{16\pi }\sqrt{-g}\left[
R +4\pi(1-\alpha)\delta_\Sigma - F_{\mu\nu}F^{\mu\nu}
\right]\no\\
&=&\mathcal{L}_{\mathcal{M}_{\text{bulk}}}+\mathcal{L}_{\mathcal{M}_{\text{singular}}}.
\end{eqnarray}

Examing only $H_{\mathcal{M}_{\text{singular}}}$, we separate the terms according to the $\delta_\Sigma$,
\begin{eqnarray}
H_{\mathcal{M}_{\text{singular}}}&=&-\frac{(1-\alpha)}{4\beta}\int_\Sigma d\Sigma\sqrt{|g|}.
\end{eqnarray}

We examine the singular set $\Sigma$ in which the cone at $r=r_H$ is made up of the surface $[t,r]$ rotating in $\Sigma$, at this point the spatial parameters are fixed and therefore do not contribute to the integral. Thus only the time parameter $t$, the zenith angle $\theta$ and the azimuth angle $\phi$ contribute, with the coordinates describing the tangent being
\begin{eqnarray}
n_1^\mu = (0,\sqrt{g^{11}},0,0),
\end{eqnarray}
Hence
\begin{eqnarray}
H_{\text{singular}}&=&-\frac{(1-\alpha)}{4\beta}\int_0^\beta \int_0^\pi \int_0^{2\pi}\left.\sqrt{|g_{22}\left(g_{00}g_{33}-g_{30}g_{03}\right)|}\right|_{\Delta_r = 0}dt d\theta d\phi = 0.
\end{eqnarray}

Deriving the free energy of the black hole state,
\begin{eqnarray}
H_{bh}=F_{bh}&=&H_{bulk}+H_{singular}\no\\
&=&\frac{1}{4\Xi l^2}\left[
-r_H^3+\Xi l^2 r_H + \frac{a^2l^2}{r_H} + \frac{q^2 l^2}{r_H}  - 2\frac{q^2 l^2}{a^2+r_H^2}r_H
\right].
\end{eqnarray}

We consider $M$ such that $T<T_C$. Due to the general properties of the thermal soliton, we examine the surface $\Sigma$ in which the conical singularity at $r=0$ is composed of tangent $r$, because at this point spacetime represented by the metric is no longer curved, with $r=0$ being definite, therefore, the spatial parameter $r$ does not contribute. It follows that the parameters that still contribute to the integral on the surface $\Sigma$ include: time $t$, zenith angle $\theta$, azimuth angle $\phi$, with the coordinates describing the tangent at the singularity being
\begin{eqnarray}
n_1^\mu = (0,\sqrt{g^{11}},0,0),\hspace{10mm}
\end{eqnarray}

the integral of the singularity has the form
\begin{eqnarray}
H_{\text{singular}}^{soliton}&=&
-\frac{(1-\alpha)}{4\beta}\int_0^\beta \int_0^\pi \int_0^{2\pi}\left.\sqrt{|g_{22}\left(g_{00}g_{33}-g_{30}g_{03}\right)|}\right|_{r = 0}dt d\theta d\phi\no\\
&=&-\frac{(1-\alpha)}{4\beta}\sqrt{a^2\left(a^2+q^2\right)}\int_0^\beta dt \int_0^\pi \cos\theta\sin\theta d\theta \int_0^{2\pi}d\phi = 0.
\end{eqnarray}

Deriving the thermal soliton state free energy,
\begin{eqnarray}
H_{soliton}=F_{soliton}&=&H_{bulk}+H_{\text{singular}}^{soliton}\no\\
&=&\frac{1}{4\Xi l^2}\left[
-r_H^3+\Xi l^2 r_H + \frac{a^2l^2}{r_H} + \frac{q^2 l^2}{r_H}  - 2\frac{q^2 l^2}{a^2+r_H^2}r_H
\right].
\end{eqnarray}

\begin{figure}[!htb]
\centering
\includegraphics[width=\textwidth]{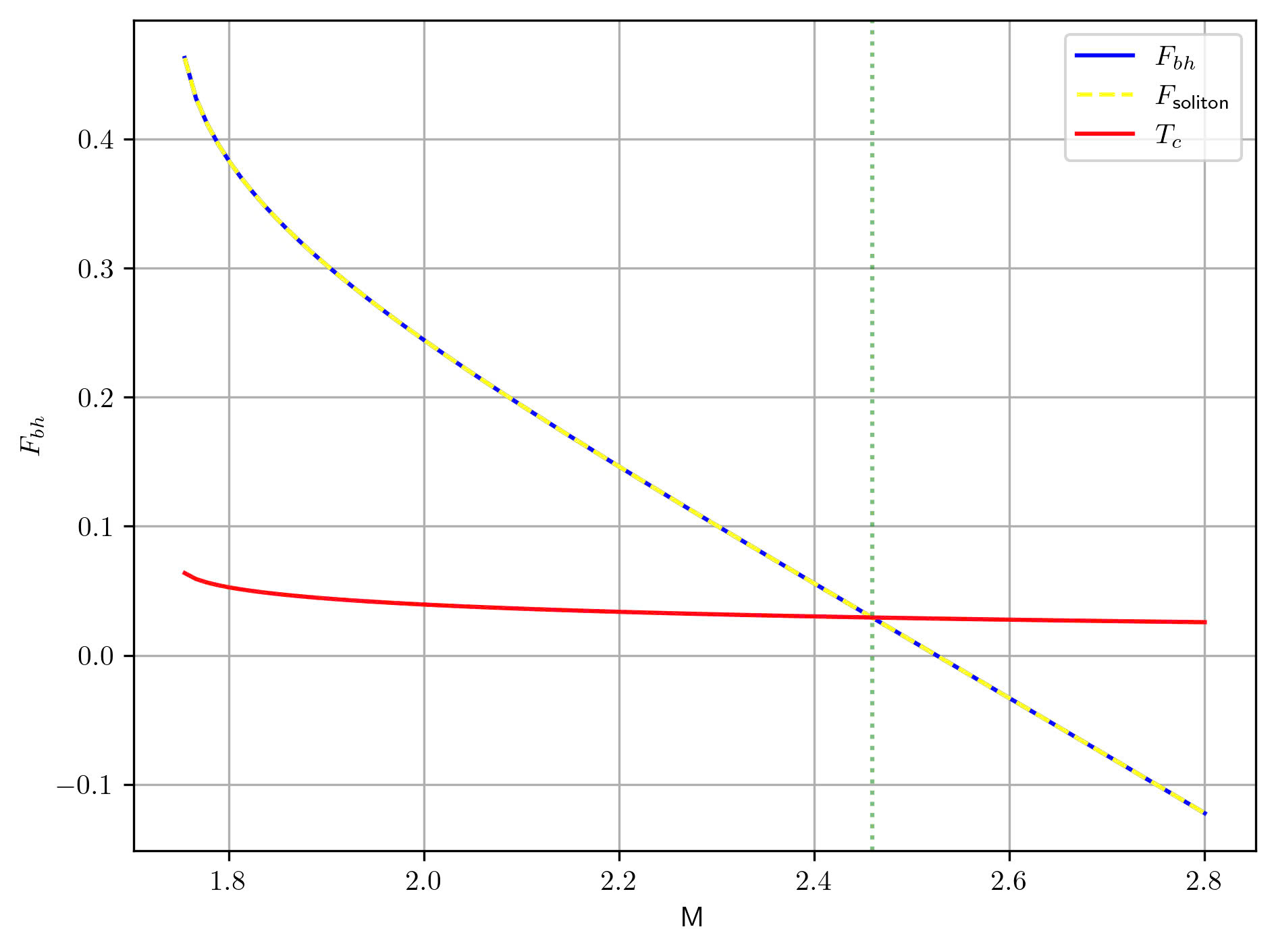}
\caption{The off-shell free energy and critical temperature at a fixed charge and the rotation of the KN black hole, in terms of $M$.}\label{6}
\end{figure}

Notably, the free energy of the black hole in Fig.\ref{6} is identical to that in Fig.\ref{5}. This arises from the rotational motion, which cancels out all the arbitrary surface contributions. Consequently, the Hamiltonian is conserved regardless of on-shell or off-shell configurations. Moreover, the free energy of the thermal soliton is also identical to that of the black hole, due to the presence of rotation. Additionally, the rotational motion removes the upper bound on the thermal soliton free energy, allowing both thermal soliton and black hole states to be considered throughout the entire thermodynamic process. Thus, the phase transition only changes the thermodynamic dominance between the thermal soliton state and the black hole state, without a distinct separation, allowing it to be interpreted as a continuous phase transition.

\subsection{Evaluation}

In the on-shell configuration, the phase transition occurs at the critical temperature, $T_C$ (Eq.\ref{kntc}), at which the transition between the thermal soliton and black hole phases takes place. The presence of electric charge imposes a minimum mass threshold for the existence of the event horizon, while rotation removes the upper mass bound of the thermal soliton state. The abrupt change in the dominant phase at the critical temperature therefore indicates a first-order phase transition. The resulting first-order phase transition agrees with that obtained from the Euclidean action, further supporting the Hamiltonian formalism as an alternative framework for studying Hawking-Page phase transitions in more general black-hole backgrounds.

Similar to the RN black hole, our off-shell analysis also reveals a continuous transition between the thermal soliton and the black hole phases. The electric charge imposes a lower mass bound, while the rotation removes the upper mass bound of the thermal soliton state. In addition, the rotation cancels out all the arbitrary surface contributions, thereby removing the distinction between the thermal soliton and the black hole states and allowing both states to be considered throughout a thermodynamic process. As the temperature rises, the black hole state gradually takes over as the thermodynamically favored state, whereas lowering the temperature progressively favors the thermal soliton state. This smooth change in the relative thermodynamic dominance is consistent with a second-order phase transition.

Thus, our off-shell analysis provides further insight into the quantum nature of the system, manifested through the shifting thermodynamic dominance between the thermal soliton and black hole states within the same thermodynamic ensemble. This result further extends the applicability of the Hamiltonian formalism to the off-shell thermodynamic description of more general black-hole backgrounds.

\section{CONCLUSION AND OUTLOOK}\label{co}

In this work, we have investigated the Hawking-Page phase transition using the Hamiltonian formalism proposed by P. M. Dirac. For the non-rotating Ba\~{n}ados-Teitelboim-Zanelli black hole in the New Massive Gravity model, we find that the on-shell Hamiltonian can be identified with the corresponding Helmholtz free energy. This agreement motivates extending the Hamiltonian formalism to more general black hole backgrounds.

We then extend the analysis to the Reissner-Nordstr\"{o}m and Kerr-Newman black holes. In the on-shell configurations, the Hamiltonian analysis reproduces the established first-order Hawking-Page transition, demonstrating the applicability of the approach to black holes with electric charge and angular momentum.

Moreover, our off-shell analysis reveals new features of the phase structure. For the Reissner-Nordstr\"{o}m black hole in Einstein-Maxwell theory, the electric charge imposes both lower and upper mass bounds on the thermal soliton state, restricting the range in which it remains relevant. For the Kerr-Newman black hole in Einstein-Maxwell theory, the presence of rotation removes the upper mass bound of the thermal soliton state and eliminates the distinction arising from arbitrary surface contributions between the thermal soliton and black hole states, allowing both states to remain relevant throughout the thermodynamic process. In both cases, the continuous change in thermodynamic dominance is consistent with a second-order phase transition and may reflect the quantum nature of the system.

Overall, our results demonstrate that the Hamiltonian formalism provides an alternative and complementary framework for investigating the Hawking-Page phase transition, reproducing the known on-shell phase structure while revealing new features in the off-shell configurations of charged and rotating black holes. The explicit construction of the Hamiltonian in these gravitational systems may also provide a useful starting point for further exploring the quantum nature of gravity, as the Hamiltonian formulation offers a natural framework for investigating the quantization of the gravitational field.

Finally, beyond the Hawking-Page transition, the Hamiltonian formalism may offer a broader framework for studying black hole phase transitions. In the present work, we have considered a fixed cosmological constant and have not yet incorporated the extended black hole thermodynamics framework, in which the cosmological constant is treated as a thermodynamic pressure. Extending the present analysis to this framework may provide a natural direction for further exploring the applicability of the Hamiltonian formalism to a broader range of black hole phase transitions. The successful applications to the non-rotating Ba\~{n}ados-Teitelboim-Zanelli, Reissner-Nordstr\"{o}m and Kerr-Newman black holes demonstrate that the formulation remains applicable to black hole backgrounds with electric charge and angular momentum. This suggests possible extensions to other types of black hole phase transitions, such as van der Waals-like, reentrant and superfluid-like phase transitions. A systematic investigation of these transitions, particularly within the extended black hole thermodynamics framework, is beyond the scope of the present work and will be considered in future studies.

\section*{ACKNOWLEDGMENTS}

This research is funded by University of Science, VNU-HCM under grant number T2025-37.

\end{document}